\documentclass[fleqn,usenatbib]{mnras}
\usepackage{newtxtext}

\usepackage[T1]{fontenc}
\usepackage{ae,aecompl}
\usepackage{threeparttable}
\usepackage{graphicx,epsfig}	
\usepackage{amsmath}	
\usepackage{amssymb}	
\usepackage{booktabs} 
\usepackage[dvipS/Names]{xcolor}
\usepackage{url}

\newcommand{\ppxf}{\textsc{pPXF }}
\newcommand{\alphaFe}{$\left[ \alpha / \text{Fe}\right]$ }

\title[Bridging the gap of compact galaxies]{Bridging the gap in the mass-size relation of compact galaxies with MaNGA}

\author[Grèbol-Tomàs et al.]{P. Grèbol-Tomàs$^{1, 2}$\thanks{E-mail: grebol@ice.csic.es}, A. Ferré-Mateu$^{2, 3}$, H. Domínguez-Sánchez$^{4}$ \\
$^{1}$ Institut de Ciències de l'Espai (ICE, CSIC), Campus UAB, Carrer de Magrans s/n, 08193 Barcelona, Spain\\
$^{2}$ Departamento de Astrofísica, Universidad de La Laguna, E-38205 La Laguna, Tenerife, Spain\\
$^{3}$ Instituto de Astrofísica de Canarias, Vía Láctea s/n, E-38205 La Laguna, Tenerife, Spain\\
$^{4}$ Centro de Estudios de Física del Cosmos de Aragón (CEFCA), Plaza San Juan, 1, 44001, Teruel, Spain}

\date{Accepted 2023 September 20. Received 2023 September 15; in original form 2023 April 22}

\pubyear{2023}

\begin{document}
\label{firstpage}
\pagerange{\pageref{firstpage}--\pageref{lastpage}}
\maketitle

\begin{abstract}

We present the analysis of the full MaNGA DR17 sample to characterize its population of compact galaxies. We focus on galaxies that fill the stellar mass (M$_{\star}$) gap between compact elliptical galaxies (cEs; $8 \lesssim \log \left(M_{\star} / M_{\odot} \right) \lesssim 10$) and compact massive galaxies (CMGs; $10 \lesssim \log \left(M_{\star} / M_{\odot} \right)$). We study their stellar populations and kinematics to reveal how their properties depend on stellar mass. We select compact galaxies in the MaNGA DR17 sample according to their effective radius ($R_e$) and stellar mass. 37 galaxies fulfill our selection criteria in the bridging region between cEs and CMGs. We derive their kinematics and stellar population parameters from the stacked spectra at 1~$R_e$ using a full spectral fitting routine. We then classify the selected compact galaxies in three main groups based on their stellar population properties. One of the groups shows characteristics compatible with relic galaxies, i.e. galaxies that have remained mostly unchanged since their early formation epoch ($z \sim 2$). Another group shows more extended and continuous star formation histories (SFHs). The third group shows a low star-forming rate at initial times, which increases at around $\sim4$ Gyr. We compare the derived properties of the selected galaxies with those of previously studied compact galaxies at different mass ranges. The selected galaxies successfully fill the mass gap between cEs and CMGs. Their properties are compatible with the assumption that the scaling relations of compact galaxies at different mass ranges are related, although galaxies in the first group are clear outliers in the fundamental plane, suggesting different formation mechanisms for this relic population. \end{abstract}

\begin{keywords}
galaxies: evolution -- galaxies: formation -- galaxies: kinematics and dynamics -- galaxies: stellar content -- galaxies: compact galaxies \\
\end{keywords}

\section{Introduction}\label{sec: Introduction}
It is well established that massive galaxies were more compact in the early Universe \citep[e.g.][]{Daddi05, Trujillo07, Buitrago08}. Although compact galaxies are observed in the local Universe, their number density increases with redshift. The realm of compact galaxies, i.e. galaxies which have smaller radii than the majority of the galaxies at a given mass, covers approximately 5 orders of magnitude in the stellar mass range. At the lowest stellar masses, $6 < \log\left( M_{\star} / M_{\odot}\right) < 8$, ultra compact dwarf galaxies (UCDs) present the smallest projected effective radii (up to $R_e \sim 20$~pc), making them the most compact galaxies in the Universe \citep[e.g.][]{Drinkwater00, Phillipps01, Brodie11}. In the intermediate mass range ($8 < \log\left(M_{\star}/M_{\odot}\right) < 10$), compact elliptical galaxies (cEs) have sizes of $100 < R_e\;\text{(pc)}<900$ \citep[e.g.][]{Faber73, Choi02, Drinkwater03}. 
Finally, the massive end of the compact realm is populated by compact massive galaxies (CMGs). These galaxies present high stellar masses ($\log \left( M_{\star}/M_{\odot} \right) > 10$) and small radii ($R_e < 1.5$ kpc ; \citealt{Shen03}), and have been extensively shown to be outliers of the local mass-size relations. 

The current galaxy formation paradigm states that the ETGs that we observe today grow in a two-phase formation scenario \citep{Bezanson09, Naab09, Oser10, Oser12, Hilz13}. In the first phase, which takes place at the earliest stages of the Universe, a gas-rich star-forming system is created \citep{Dekel09a}. The result is an extremely compact object, often referred to as a \textit{blue nugget}. These galaxies show blue colors and high luminosity \citep{Zolotov15}, fuelled by an intense star formation (SFR$\geq 10^3 \text{M}_{\odot} \; \text{yr}^{-1}$; \citealt{Smith20}). At some stage this dissipative phase ends, and the compact object is quenched, becoming a massive, red and metal-rich object. These CMGs, also nick-named \textit{red nuggets} \citep{Damjanov14, Schreiber18, MartinNavarro19, Valentino20}, mark the end of the first phase of formation, at most by $z\sim 2$. Recent observations with the James Webb Space Telescope have discovered red nuggets even at $z \sim 7$ \citep{Nanayakkara2022, Carnall23a, Carnall23b}. 

The second phase of ETGs growth is driven by dry minor merger events, which induce the growth of the red nugget, by adding accreted material to the outskirts of these galaxies. This process could explain the mild grow in stellar mass but the large increase in size, driving the strong size evolution seen over cosmic time and building up the massive ETGs population observed at $z=0$ \citep{Daddi05, Trujillo07, vanDokkum10}. 

Since the second phase is driven by stochastic events, there is a low probability that a galaxy avoids such a phase, remaining unchanged since the early stages and presenting the properties of a red nugget \citep{Trujillo09b, Quilis13, Poggianti13b, Damjanov14}. These untouched galaxies, found at $z~\sim~0$, are often referred to as \textit{massive relic galaxies}. These are rare and hard-to-find objects but extremely valuable for understanding galaxy formation, as they have similar properties as high-$z$ ETGs but observed with the spectral and spatial resolution of local galaxies. Their importance relies on their relation with those of ETGs at high redshift \citep[][]{Gargiulo16, Belli17, Tanaka19}. Theoretical models predicting red nuggets survival are sensitive to galaxy merging processes \citep{Wellons16}, and their estimates expect that 0.15\% of the massive galaxies formed at $z\simeq 2$ could end up being a massive relic galaxy \citep{Quilis13}. 

The current number of confirmed massive relic galaxies up to $z\sim 0.5$ is 13 \citep{Trujillo14, FerreMateu17, Spiniello21a}. However, only three massive relic galaxies in the local Universe have been characterized in full detail: Mrk1216, PGC032873, NGC1277 \citep{Trujillo14, FerreMateu17}. All these massive relic galaxies feature a high mass ($\log \left( M_{\star} / M_{\odot} \right) > 11$) and small radius ($R_e < 1$ kpc), with a fast star formation episode as early as at the time of the Big Bang ($t \sim t_{BB}\sim 14$~Gyr). They all present disk-like morphologies, similar to those in observed massive red nuggets \citep{Buitrago08, vanderWel11, Trujillo14}. 

Massive relics are found in all environments, although they seem to thrive in clusters of galaxies \citep{Poggianti13b, Cebrian14, Damjanov15env, Stringer15, PeraltadeArriba16}. This is a combination of serendipitiousness (e.g. finding them in a cluster is easier), and the extreme conditions from the cluster itself. The high gravitational fields accelerate the galaxies, and given their high velocities, this prevents mergers from taking place, promoting the occurrence of massive relic galaxies. Conditions in the field are expected to happen later and therefore galaxies in this environment will tend to be less extreme in their properties \citep{FerreMateu17}. In the latter, a `degree of relicness' linked to the environment was proposed for the known massive relic galaxies, later supported by \cite{Spiniello21b}.

However, not all CMGs are massive relic galaxies. In fact, the majority of CMGs found in the local Universe show surprisingly large fractions of young stellar populations (e.g. \citealt{Trujillo09a}, \citealt{Poggianti13}, \citealt{Damjanov14}, \citealt{Buitrago18}). 
How these galaxies are formed still poses a great challenge within the current cosmological models \citep{FerreMateu12}. While some of them could be the remnant of a more massive galaxy that has lost its stars due to external processes, such formation scenario is less likely to happen and CMGs are mostly expected to be formed by in-situ processes \citep{Cappellari16}.

As we move towards lower stellar masses, the leading formation scenario changes from an in-situ based to external processes playing a more relevant role. This change seems to occur around the characteristic mass scale of 3$\times10^{10}M_{\odot}$ \citep[e.g.][]{Cappellari16, FerreMateu18, FerreMateu21b, DominguezSanchez20}, where several relations of ETGs seem to have relevant changes. As a result, cEs are thought to be a mix-bag of objects, although they are mostly thought to be the result of stripping a dwarf elliptical galaxy or a low-mass ETG or spiral \citep[e.g.][]{Faber73, Bekki01, Choi02, Graham02, Paudel16}. However, some of these galaxies are also expected to form in situ, describing cEs like the true low-mass end of ETGs (e.g. \citealt{Kormendy09a, Kormendy12, Du19}). 

In the first case, where the cE is the result of a stripping process, they are expected to be outliers in most of the scaling relations, such as the black hole-galaxy mass, or the mass-metallicity relation \citep[e.g.][]{Norris14, Janz16, FerreMateu18, FerreMateu21b, Kim20}. This is further supported by the large SMBHs typically found in their centers \citep[e.g.][]{Forbes14, Paudel16, Pechetti17, FerreMateu21b}. But the strongest evidence for this evolutionary path has been seen observationally, with cEs currently being stripped by a larger galaxy (e.g. \citealt{Huxor11a, Paudel14a, FerreMateu18}). Nonetheless, evidence for some cEs being formed in-situ has also been seen, in particular outside the cluster environment, where stripping is not likely to happen \citep{Huxor13, Paudel14b, FerreMateu18, FerreMateu21a, Kim20}. As they are thought to be the very low-mass end of ETGs, it is expected that they will follow the scaling relations at such low-mass end.

Unfortunately, there is no precise number of compact galaxies for each formation pathway yet (e.g. in-situ vs. ex-situ), due to the incomplete samples we have at hand. Interestingly, \cite{FerreMateu21b} suggested that there may be a connection between the cEs and CMGs families. In their mass-size relation plot \citep[][Figure 11]{FerreMateu21b}, they showed that the distribution of CMGs and cEs presented similar stellar populations, while also sharing similar kinematic features. However, there was a noticeable gap between these two groups of compact galaxies, which could be the clue to reveal whether such connection exists in reality. To this end, the following study is aimed at looking for compact galaxies bridging this gap. We study their kinematic and stellar populations properties, in order to relate the evolutionary paths of compact galaxies at different masses. To that purpose, we use local galaxies from the MaNGA survey \citep{Bundy15} due to their large statistics and wealth of data, including IFU observations. In this work we present the study of the MaNGA sample and the global properties of the selected compact galaxies, whereas the spatially-resolved analysis will be done in a future work. 

In Section~\ref{sec: Sample} we present the MaNGA survey and our criteria to select compact galaxies. In Section~\ref{sec: analysis} we obtain the main kinematic and stellar populations properties of the selected sample. We then classify the compact galaxies in different groups based on these properties. In Section~\ref{sec: discussion} we discuss the stellar population and kinematic properties of each group independently and we compare them with the properties shown by cEs and CMGs in the literature. Finally, we present in Section~\ref{sec: summary} a summary of our conclusions by sketching how the properties of each group can be linked with different galaxy formation pathways.

\section{Sample}\label{sec: Sample}
In this work, we use the Mapping Nearby Galaxies at APO (MaNGA; \citealt{Bundy15}) survey, a Sloan Digital Sky Survey (SDSS; \citealt{York00}) survey. With its latest data release, DR17, spectroscopy for over 10 000 galaxies up to $z<0.17$ has been obtained. This survey takes advantage of the \textit{Integrated Field Unit} (IFU) technology to obtain spatially resolved spectroscopy for each single galaxy. Data is presented as datacubes, where two dimensions correspond to spatial coordinates (known as \textit{spaxels}) and the third contains the spectrum of each spaxel. The spectra cover a wavelength range from 3600 \AA\,  to 10300 \AA\, with a spectral resolution of $R\sim 2000$, which roughly corresponds to 2.51~\AA\,  at 5000 \AA. 

We select compact galaxies by imposing mass and size criteria. The structural photometric parameters are obtained from the MaNGA PyMorph DR17 catalog \citep{Fischer19}. It provides parameters from fitted Sérsic and Sérsic+Exponential profiles to the 2D surface brightness profiles of MaNGA DR17 galaxies. From this catalog we use the effective radii ($R_e$), axis ratios and galaxy's luminosities. The latter are translated into stellar mass ($M_{\star}$) using the mass-to-light ratio from \cite{Mendel14}. The MaNGA PyMorph catalog provides a flagging system (FLAG\_FIT) which separates galaxies which are better described by a Sérsic or a Sérsic+Exponential profile. We therefore use the parameters returned by the optimal model for each galaxy. When FLAG\_FIT equals 0 (both models are acceptable), we use the parameters returned by the Sérsic+Exponential parametrization.

\begin{figure*}
\centering
\includegraphics[width=1\textwidth]{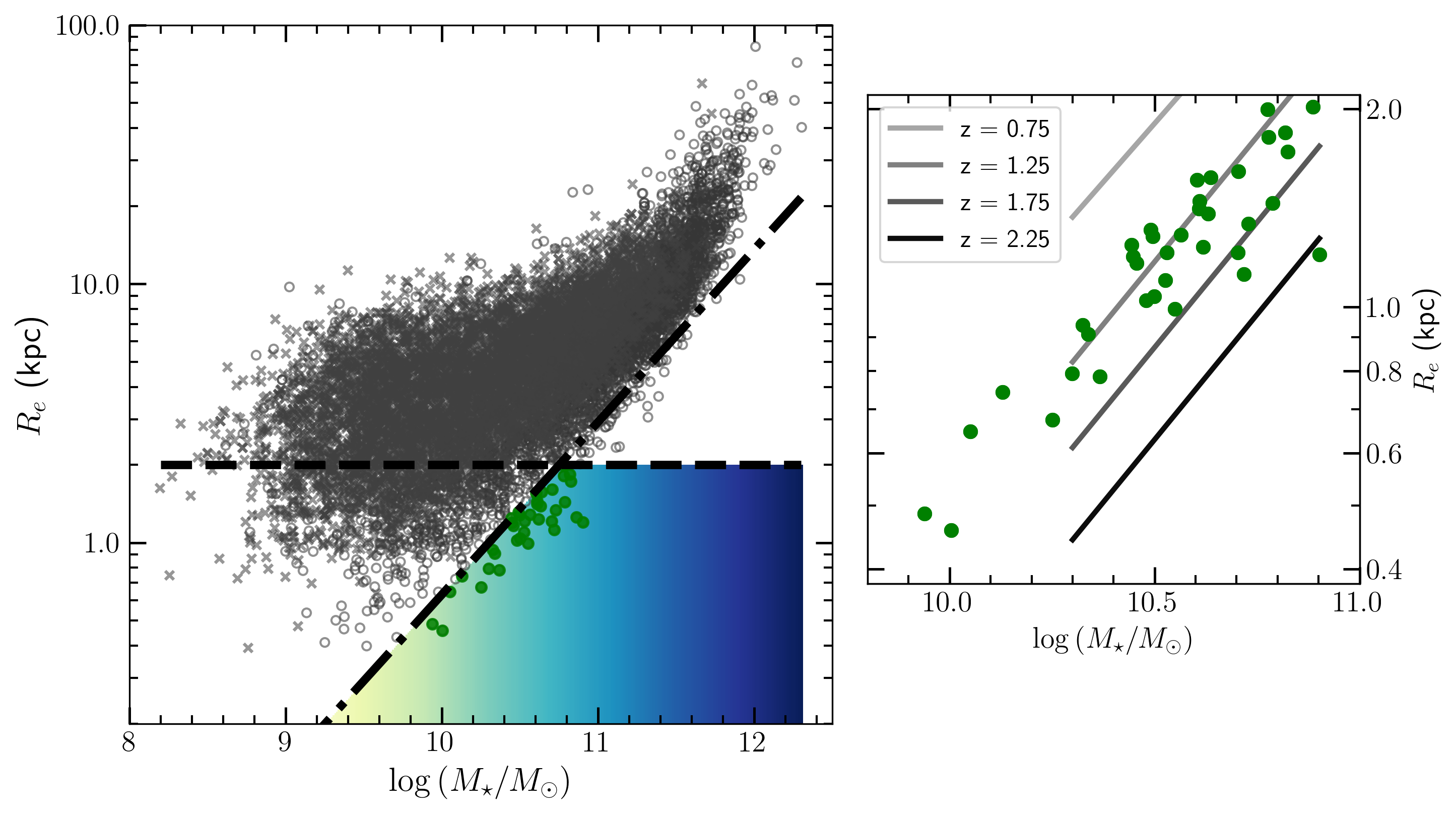}
\caption{$R_e$ vs. $M_{\star}$ for the full MaNGA DR17 dataset (10293 nearby galaxies). The effective radii are extracted from MaNGA's PyMorph catalog \citep{DominguezSanchez22}. Mass values are obtained by applying the mass-to-light ratios from \citet{Mendel14} to the luminosities presented in PyMorph. We use the MaNGA Deep Learning Morphological Value Added catalog \citep{DominguezSanchez22} to classify galaxies according to their morphology. Crosses indicate spiral or $S0$ galaxies, while circles represent elliptical galaxies. The dash-dotted black line represents the compactness criterion defined in Section \ref{sec: Sample}, based on the conditions set in \citet{Barro13, Charbonnier17}. The black dashed line marks the upper threshold of $R_e = 2$ kpc. The background colors intuitively show the compact galaxy family in terms of stellar mass. Yellowish colors represent the high-mass end of the cEs family, whereas the blue region represents the CMG region. In the inset zoom figure we overplot different mass-size relations for ETG at various redshifts from \citet{vanderWel14}, along with the selected compact galaxies for this work.}
\label{fig:Mass radius}
\end{figure*}

There are a number of different criteria in the literature to define compact galaxies, particularly at the high-mass end \citep[e.g.][]{Buitrago18, Valentinuzzi10, Scognamiglio20}. Since we are aiming to fill the gap that connects the high-mass end of cEs with the low-mass end of CMGs, here we impose the following criteria:
\begin{itemize}
    \item $M_{\star} > 10^9 \; M_{\odot}$
    \item $R_e < 2 \text{ kpc}$
    \item $\log \left( \Sigma_{1.5} \right) > 10.3 \text{ dex}$
\end{itemize}

where $\Sigma_{1.5} = \frac{M\; [M_{\odot}]}{(R_e\; [\text{kpc}])^{1.5}}$ is a modified surface mass density, as in \citet[][]{Barro13}. The first condition is set to select all galaxies with stellar masses that cover the high-mass end of cEs, which is sometimes mixed with the low-mass end of regular elliptical galaxies \citep[see e.g.][Figure 11]{FerreMateu21b}. The second corresponds to the largest size limit used to select CMGs \citep[][]{Buitrago18, Charbonnier17}. The third criteria ensures the compactness of the candidate (based in \citealt{Barro13, Damjanov15ndens, Charbonnier17}). 

We show in Figure \ref{fig:Mass radius} the $R_e$ vs M$_{\star}$ for the complete sample of 10293 MaNGA DR17 galaxies. Galxies are marked according to their morphology. We take advantage of the morphological classification presented in the MaNGA Deep Learning Morphological Value Added catalog \citet[MDLM-VAC]{DominguezSanchez22}. It provides a series of binary classifications which separate ETGs from LTGs, pure ellipticals (Es) from lenticulars (S0), barred from non-barred galaxies and edge-on from non-edge-on galaxies. In addition, the catalog also reports a T-Type value, analogue to the \cite{Hubble1926} sequence. For the figure, we select ETGs by requiring: \texttt{T-Type<0} and \texttt{PS0<0.5}. The selection results in 3834 out of 10293 galaxies from the MaNGA DR17 parent sample.

In Figure \ref{fig:Mass radius},  compact galaxies selected by the above criteria are shown in green, being all of them ellipitcals. This selection returns 38 galaxies. After visually inspecting each individually, we discard object 8092~-~12794 as it corresponds to two interacting galaxies. Our final sample thus consists of 37 compact galaxies. Their stellar masses, effective radii and redshifts are quoted in Table \ref{tab:Selected compact massive galaxies}. The region corresponding to where compact galaxies would be located is colored in Figure \ref{fig:Mass radius}. The yellowish region shows the high-mass end of cEs, while CMGs are expected to populate the blueish region. The galaxies selected in this work do, precisely, fall in the mass gap between cEs and CMGs (turquoise region). These mass limits are purely illustrative, as there is not a unique mass threshold in the literature to distinguish compact families.

In fact, uncertainties in the $M_{\star}$ and $R_e$ estimations can vary the number of selected galaxies. To check the robusness of our selection, for each galaxy, we have considered the combination of mass and $R_e$ more favourable and more difabourable, within errors, to be considered as compact. We assume a  standard error on the $M_{\star}$ of 20\%, while the error in Re is quoted from the PyMorph catalog.

The most favorable conditions (largest $M_{\star}$, smallest $R_e$) would provide 63 galaxies, an increase of $\times$1.5 in respect to the nominal value. The least favorable ones (smallest $M_{\star}$ and largest $R_e$) would instead only provide 23 galaxies, roughly 60\% of the selection from the nominal values. This galaxies are therefore considered as the most robust (highlighted in Table \ref{tab:Selected compact massive galaxies}), but we will use hereafter the nominal values.

\begin{table}
\centering
\caption{The 37 selected MaNGA compact galaxies. Each galaxy is labelled according to its Plate-IFU given by the MaNGA DR17 survey. The redshift value is obtained from the NASA-Sloan Atlas catalog\tnote{1}. Stellar mass and effective radius values (including their errors) are estimations from the PyMorph and Deep Learning VACs \citep[][]{DominguezSanchez22, Fischer19}. Stellar mass errors are asumed to be uniform and the 20\% of the nominal $M_{\star}$ value. Galaxies with an asterisk in their Plate-IFU are those selected even when considering the most unfavorable $M_{\star}$ and $R_e$ values according to their uncertainties.}
\label{tab:Selected compact massive galaxies}
\begin{threeparttable}
\begin{tabular}{lccc}
\toprule
\textbf{Plate-IFU} & $\boldsymbol{\log\left( M_{\star} / M_{\odot}\right)}$ & $\boldsymbol{R_e}$ \textbf{[kpc]} & $\boldsymbol{z}$ \\ \midrule
8443-1901*  & 10.61 & 1.41 $\pm$ 0.01 & 0.0294 \\
8721-1901*  & 10.55 & 0.99 $\pm$ 0.01 & 0.0456 \\
8323-1901  & 10.61 & 1.45 $\pm$ 0.01 & 0.0253 \\
9492-1901  & 10.46 & 1.16 $\pm$ 0.02 & 0.0457 \\
9042-1901  & 10.78 & 1.81 $\pm$ 0.29 & 0.0729 \\
7977-1901  & 10.05 & 0.65 $\pm$ 0.01 & 0.0260 \\
9869-1901  & 10.82 & 1.84 $\pm$ 0.33 & 0.0309 \\
8601-1902  & 10.44 & 1.24 $\pm$ 0.15 & 0.0280 \\
9507-1902  & 10.82 & 1.72 $\pm$ 0.28 & 0.0517 \\
8243-1902*  & 10.53 & 1.21 $\pm$ 0.02 & 0.0427 \\
8616-1902  & 10.78 & 1.99 $\pm$ 0.02 & 0.0302 \\
8486-1902*  & 10.34 & 0.91 $\pm$ 0.02 & 0.0195 \\
8464-3703*  & 10.30 & 0.79 $\pm$ 0.01 & 0.0398 \\
8137-3703  & 10.49 & 1.31 $\pm$ 0.009 & 0.0317 \\
10216-3703* & 10.63 & 1.39 $\pm$ 0.02 & 0.0466 \\
10510-1901* & 10.00 & 0.46 $\pm$ 0.01 & 0.0197 \\
11011-1901 & 10.49 & 1.28 $\pm$ 0.007 & 0.0270 \\
11020-1902* & 10.90 & 1.20 $\pm$ 0.01 & 0.0253 \\
11827-1901* & 10.73 & 1.34 $\pm$ 0.01 & 0.0265 \\
11943-9102* & 10.50 & 1.04 $\pm$ 0.01 & 0.0284 \\
11945-1901* & 10.72 & 1.12 $\pm$ 0.03 & 0.0292 \\
11979-1902* & 10.70 & 1.61 $\pm$ 0.03 & 0.0469 \\
11984-1902* & 10.37 & 0.78 $\pm$ 0.01 & 0.0464 \\
8248-3704*  & 10.70 & 1.21 $\pm$ 0.01 & 0.0255 \\
8710-1902*  & 9.94  & 0.48 $\pm$ 0.01 & 0.0214 \\
9496-1902*  & 10.48 & 1.02 $\pm$ 0.01 & 0.0452 \\
9880-1902  & 10.13 & 0.74 $\pm$ 0.01 & 0.0255 \\
11954-1902 & 10.45 & 1.19 $\pm$ 0.01 & 0.0266 \\
12067-3701* & 10.25 & 0.67 $\pm$ 0.01 & 0.0379 \\
7981-1902*  & 10.79 & 1.44 $\pm$ 0.02 & 0.0300 \\
11015-1902* & 10.56 & 1.29 $\pm$ 0.03 & 0.0241 \\
11960-1902* & 10.89 & 2.01 $\pm$ 0.01 & 0.0514 \\
11967-3702* & 10.62 & 1.23 $\pm$ 0.02 & 0.0458 \\
11968-1902 & 10.60 & 1.56 $\pm$ 0.01 & 0.0451 \\
11969-1902 & 10.64 & 1.57 $\pm$ 0.01 & 0.0493 \\
12624-3702 & 10.32 & 0.94 $\pm$ 0.02 & 0.0272 \\
12673-1901* & 10.53 & 1.10 $\pm$ 0.01 & 0.0274 \\ \bottomrule
\end{tabular}%
\begin{tablenotes}
\item[1] \url{https://www.sdss4.org/dr17/manga/manga-target-selection/nsa/}
\end{tablenotes}
\end{threeparttable}
\end{table}

According to the two-phase formation paradigm, compact galaxies should be already formed by at least $z\sim 2$, after the red nugget is formed. If compact galaxies in the mass gap are somehow the remnants of this early stage or directly connected to it, they should roughly match the mass-size relation at $z\sim 2$. For example, all massive relic galaxies studied to date are consistent with the mass-relation of $z\sim 2$ galaxies \citep[e.g.][]{FerreMateu17, Yildirim17, Spiniello21a}. In the right panel of Figure \ref{fig:Mass radius} we compare the 37 selected compact galaxies to mass-size relations at different redshifts using CANDELS/3D-HST (from \citealt{vanderWel14}). We find that, although all the galaxies in our sample are found in the nearby Universe, they are in reality more compatible with the mass-size relations at $z\sim 1.25-1.75$. Only the most massive galaxy, 11020-1902, is compatible with a $z\sim$2 relation, making it the best candidate for a relic galaxy in this sample. We will discuss this particular galaxy in more detail in Section \ref{subsec: stellar pops}.

\subsection{Methodology}\label{sec: methodology}
In this work we aim at characterizing the global properties of the galaxies bridging the gap between cEs and CMGs in the mass-size relation of the MaNGA DR17 sample. In order to increase the signal-to-noise ratio and to simplify the statistical analysis, we have stacked together all the spaxels within 1~$R_e$ for each galaxy cube. We have used a \texttt{.dpuser} code applied to the QFitsView FITS file viewer \citep{Ott12} to stack the spectra. For each galaxy, we have retrieved a single spectrum from stacking all pixels within 1~$R_e$ and the effective radius value from PyMorph VAC. We have centered the circular stacking region at the pixel with the highest photon count.

\subsection{Full spectral fitting}\label{subsubsec: full spectral fitting}
Kinematic and stellar populations from stacked galaxy spectra are obtained using the Penalized Pixel-Fitting method (\ppxf, \citealt{Cappellari04}), implemented in the \ppxf Python package \citep{Cappellari12ppxf}, and the GandALF routine \citep{Sarzi06}. We have used the full MILES stellar population models \citep{Vazdekis15} to fit the spectra, in a wavelength range between 3800 \AA\; and 5600 \AA, with a nominal spectral resolution of FWHM~=~2.5\AA \citep{FalconBarroso11}. The stellar models considered stellar ages from 0.03 to 14~Gyr and metallicities between -2.27 and +0.40~dex. We have used the Base models, corresponding to BaSTI isochrones \citep{Pietrinferni14, Hidalgo18}. 

Massive relic galaxies have shown to phave an overall steep IMF \citep[][]{MartinNavarro15, FerreMateu17, MartinNavarro23}. \cite{FerreMateu13} characterized the impact of the IMF on the derived star formation histories (SFHs). They showed that a slight change in the IMF slope does not significantly change the results of the derived SFHs. Based on their conclusions, we worked with a Kroupa-bimodal function of $\;\Gamma = 1.30$, so that we can compare our results with previous works.

\subsubsection{Stellar kinematics} \label{subsubsection: kinemtatics methodology}
\begin{figure}
    \centering
    \includegraphics[width = 0.5\textwidth]{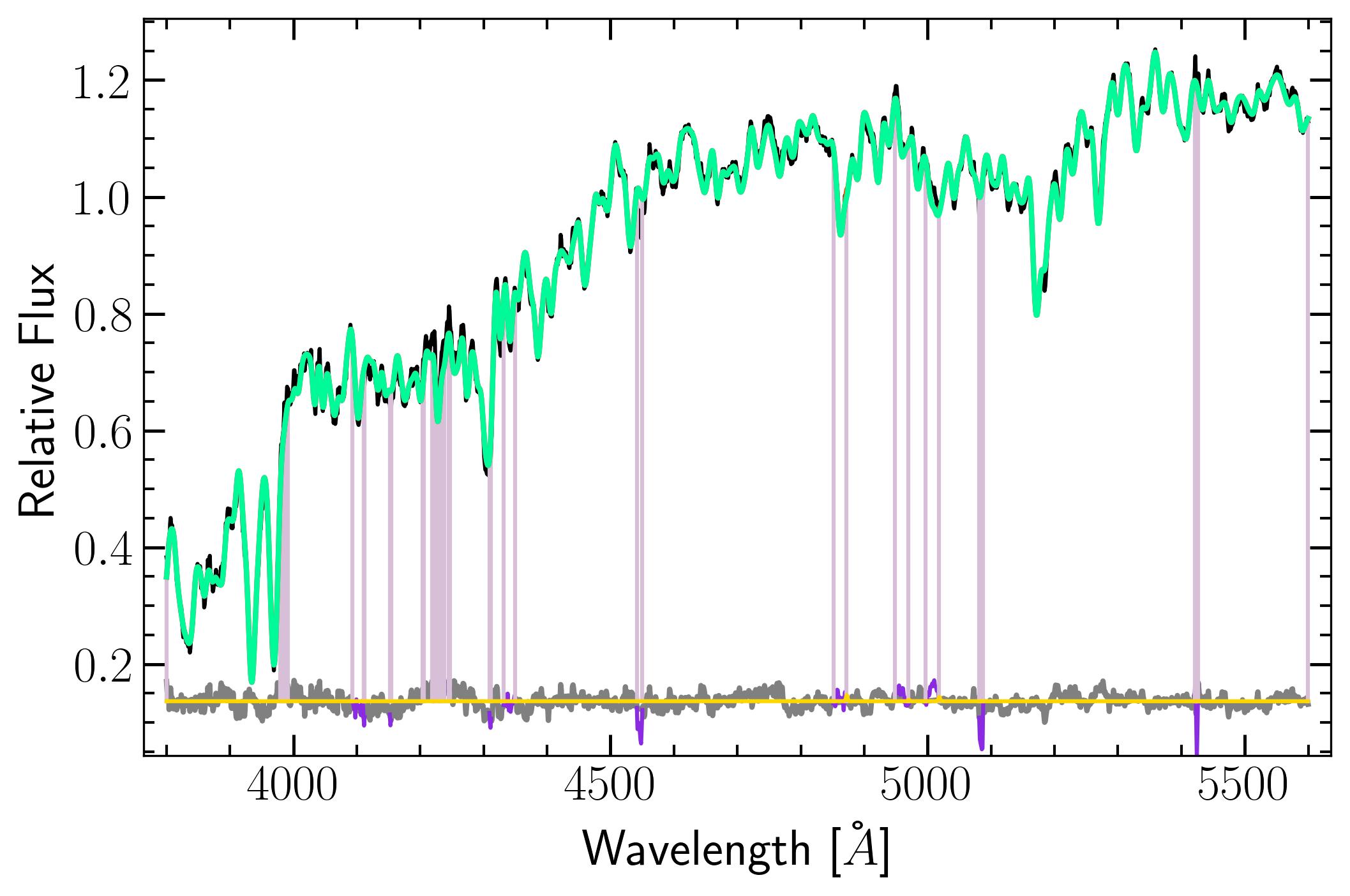}
    \caption{Spectrum of the selected compact galaxy 11943-9102 (black) and its fitted spectrum using \ppxf (green). Purple ranges correspond to masked regions in the spectrum and the gray line shows the residuals from the fit. The yellow line is the emission line result from the fitting. In this case, there is zero emission (horizontal line), which is a further evidence of passive galaxy. The fit has been derived using the full MILES stellar population models with nominal resolution FWHM~=~2.5~\AA~with Base-Fe models.}
    \label{fig: ppxffit}
\end{figure}

The stellar kinematic measurements were obtained with the \ppxf routine with an additive Legendre polynomial of degree 5 (used to correct the template continuum shape)\footnote[2]{We tested different values for the polynomia, following a similar methodology to \cite{dAgo23}. We found that 5 is the one that optimizes the fitting errors}. From this first \ppxf iteration, we derive the recessional velocity, $v$, and its velocity dispersion,~$\sigma$. These two parameters are obtained after fitting the line-of-sight velocity distribution, $\mathcal{L}\left( \mathcal{V} \right)$, as Gauss-Hermite series \citep{vanderMarelFranx93, Gerhard93}:
\begin{equation}
    \mathcal{L}\left( \mathcal{V} \right) = \frac{e^{-y^2/2}}{\sigma\sqrt{2\pi}}\left[ 1+ \sum_{m=3}^M h_m H_m\left(y\right)\right],
\end{equation}

where $y = \left( \mathcal{V} - v \right) / \sigma$ and $H_m \left(y\right)$ are Hermite polynomials. As suggested in \citet{Cappellari04} a first second-order fitting is conducted to recover $v$ and $\sigma$. A second fitting with $v$ and $\sigma$ fixed was applied to retrieve higher order kinematic Hermite coefficients. Figure \ref{fig: ppxffit} illustrates this procedure by showing the spectrum of a random galaxy in our sample fitted with the \ppxf routine.

Another relevant kinematic parameter is the specific angular momentum, $\lambda_R$ \citep[][]{Emsellem07}. It provides information about the internal dynamics of the galaxy, and it is commonly used to classify galaxies as fast or slow rotators \citep[e.g.][]{Zoldan18, Sweet20, Romeo23}. This dichotomy has been found to be related with the galaxy morphology, with most massive ETGs being more likely slow rotators \citep{FalconBarroso19}. Here we calculate the $\lambda_R$ values as in \citet{Fischer19}, using the IFU observations provided by the MaNGA survey. The $\lambda_R$ is calculated as a weighted mean over the values in each spaxel:
\begin{equation}
    \lambda_R = \frac{\sum_i^N R_i F_i \left| v_i \right|}{\sum_i^N R_i F_i \sqrt{v^2_i + \sigma_i^2}},
    \label{eq: lambda}
\end{equation}

where $R$, $F$, $v$ and $\sigma$ denote the radial position, flux, rotational velocity and velocity dispersion at the $i$-th spaxel. The sum is done up to $1~R_e$ for spaxels with $S/N > 5$. The number of spaxels used in the $\lambda_R$ estimation strongly depends on the projected angular size of the galaxy. When deriving $\lambda_R$ for our 37 compact galaxies, 13 of them did not have $S/N$ high enough to estimate this parameter. This corresponds to a 32\% of the total number of selected compact galaxies. For the other 24 galaxies for which we could calculate their $\lambda_R$, the typical value for spaxels used was $\sim 40$. In all cases, all the spaxels within $1\;R_e$ fulfilled the requirements to be included in the $\lambda_R$ calculation. Additionally, $\lambda_R$ were corrected for seeing following \citet{Graham18}. The $\lambda_R$ values are shown in Table \ref{tab: kinematics stell pops}.

\subsubsection{Stellar populations, SFHs and characteristic timescales} \label{subsubsec: SFH  characterisation}
We run \ppxf again, fixing the kinematics to the values obtained in the first iteration and using a multiplicative Legendre polynomial of degree 7 (to correct for low-frequency continuum variations). From this second run we obtain the mean stellar ages and total metallicities, but also the SFHs of each galaxy. 

\begin{figure}
    \centering
    \includegraphics[width=0.45\textwidth]{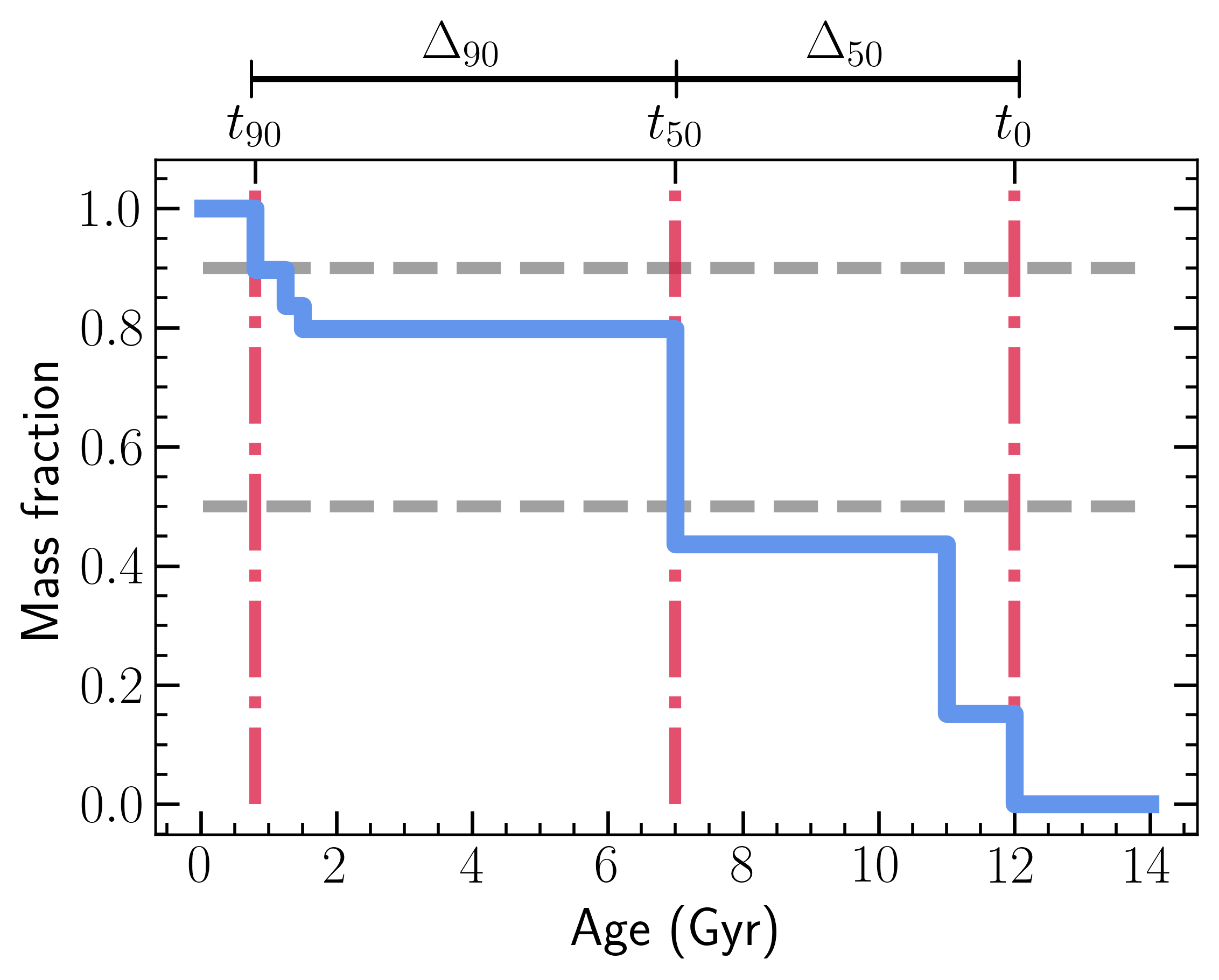}
    \caption{Example of a derived SFH using \ppxf, corresponding to the galaxy 11954-1902. The figure shows how the mass fraction of the galaxy increases with time. The steep increments are the result of not applying a regularization in the \ppxf routine. Dashed gray lines show the 50\% and 90\% values of the total mass fraction. Dash-dotted vertical lines mark the position of $t_0$, $t_{50}$ and $t_{90}$ parameters, the times at which the total mass fraction surpasses 0\%, 50\% and 90\%, respectively. Based on these parameters, we define $\Delta_{90} = t_{90} - t_{50}$ and $\Delta_{50} = t_{50} - t_0$ to characterize the SFH of a galaxy.}
    \label{fig: singleSFH}
\end{figure}

As illustration, we present in Figure \ref{fig: singleSFH} an illustrative SFH of a galaxy in our sample. In this case, the SFH is shown as the `cumulative' stellar mass that the galaxy builds up over cosmic time. From this, several characteristic look-back times can be computed, such as the time when the galaxy formed the 50\% and 90\% of its stellar mass ($t_{50}$ and $t_{90}$, respectively). We also define $t_0$ as the look-back time at which the galaxy started forming stars, which does not need to correspond to the time of the Big Bang. These look-back times are shown in Figure \ref{fig: singleSFH} as red vertical dash-dotted lines. From these times, we define the characteristic timescales that can provide information about how fast the star formation occurred. We define $\Delta_{90} = t_{90} - t_{50}$ and $\Delta_{50} = t_{50} - t_0$, also shown in Figure \ref{fig: singleSFH}. For example, a high value of these parameters is representative of an extended SFH, whereas a low value would represent very early and fast formation timescales. Table \ref{tab: kinematics stell pops} quotes the most relevant stellar population properties derived in this section. 

The stellar population parameters can be affected by the stellar population models employed \citep{DominguezSanchez19}. In addition, the use of scaled-solar or $\alpha$-enhanced models can also impact on the results, as shown in \citet{Spiniello21a, Spiniello21b, dAgo23}. The reader is referred to Appendix \ref{app: models}, where we have investigated the possible impact from the use of different SSP models. We conclude that our results are robust agaist $\alpha$-enhancements and IMF slopes (except for very steep IMF values). It is out of the scope of this paper to study the effect of using different SSP libraries.

For the determination of the $\alpha$-abundance we use the more classical line index technique. We compare the metallic absorption line strengths of Mg$_b$, and <Fe> (a combination of Fe5270, Fe5335) \citep{Gonzalez1993} for the scaled solar ([$\alpha$/Fe]=0.0\,dex) and enhanced ([$\alpha$/Fe]=0.4\,dex) SSP models. However, because this method is based on single features, it is also prone to have some of the lines affected by bad sky residuals or bad pixels in the spectra. To minimize this effect, rather than measuring this value for each galaxy individually, we will only measure the value for the three classes described in Section \ref{subsec: clustering}. 

\begin{table*}
\caption{Kinematic and stellar population results from the analysis of the stacked spectra covering 1 $R_{e}$. Stellar velocity, stellar velocity dispersion, age and metallicity are the output of the \ppxf analysis \citep[][]{Cappellari04, Cappellari12ppxf}. The rotational parameter $\lambda_R$ is extracted as in \citet{Fischer19}, which could not be calculated for a handful of galaxies. This parameter is corrected according to the PSF following \citet{Graham18}. $\Delta_{50}$ and $\Delta_{90}$ are two parameters that can be used to characterize galaxy formation timescales. The last column corresponds to the galaxy allocation in groups according to the $k$-means classification (see Section \ref{sec: analysis}).}
\label{tab: kinematics stell pops}
\resizebox{\textwidth}{!}{%
\begin{tabular}{lcccccccr}
\toprule
\textbf{Plate-IFU} & $\boldsymbol{v}$ & $\boldsymbol{\sigma}$ & $\boldsymbol{\lambda_R}$ & \textbf{Age} & \textbf{[M/H]} & $\boldsymbol{\Delta_{50}}$ & $\boldsymbol{\Delta_{90}}$ & \textbf{Cluster} \\
 & \textbf{(km s$\boldsymbol{^{-1}}$)} & \textbf{(km s$\boldsymbol{^{-1}}$)} &  & \textbf{(Gyr)} & \textbf{(dex)} & \textbf{(Gyr)} & \textbf{(Gyr)} &  \textbf{group} \\\midrule
8443-1901 & $95.40\pm0.66$ & $162.69\pm1.62$ & 0.40 & $9.03\pm2.32$ & $0.127\pm0.043$ & 3.00 & 6.50 & B \\
8721-1901 & $109.00\pm1.24$ & $206.73\pm4.28$ & - & $12.96\pm0.40$ & $0.307\pm0.017$ & 0.00 & 0.00 & A \\
8323-1901 & $84.76\pm0.67$ & $178.21\pm2.38$ & 0.77 & $10.24\pm1.31$ & $0.335\pm0.050$ & 2.00 & 0.00 & A \\
9492-1901 & $117.05\pm1.00$ & $197.28\pm2.23$ & - & $13.19\pm1.02$ & $0.149\pm0.019$ & 0.00 & 0.00 & A \\
9042-1901 & $106.08\pm1.17$ & $229.64\pm5.98$ & - & $12.79\pm1.11$ & $0.318\pm0.080$ & 0.00 & 0.00 & A \\
7977-1901 & $91.69\pm1.29$ & $158.27\pm5.52$ & - & $11.49\pm1.27$ & $0.213\pm0.043$ & 2.50 & 0.50 & A \\
9869-1901 & $99.77\pm0.85$ & $197.83\pm3.24$ & 0.19 & $13.34\pm0.62$ & $0.388\pm0.039$ & 0.00 & 0.00 & A \\
8601-1902 & $98.63\pm0.66$ & $178.83\pm5.17$ & 0.16 & $13.62\pm0.73$ & $0.223\pm0.077$ & 0.00 & 0.00 & A \\
9507-1902 & $114.46\pm1.98$ & $291.23\pm6.62$ & 0.65 & $13.32\pm0.57$ & $0.272\pm0.034$ & 0.00 & 1.00 & A \\
8243-1902 & $106.73\pm0.86$ & $231.91\pm5.98$ & - & $13.27\pm0.87$ & $0.363\pm0.034$ & 0.50 & 0.00 & A \\
8616-1902 & $89.86\pm1.47$ & $205.35\pm6.16$ & 0.67 & $11.41\pm1.90$ & $0.293\pm0.053$ & 0.00 & 3.00 & A \\
8486-1902 & $102.03\pm0.76$ & $170.74\pm4.82$ & 0.31 & $11.91\pm0.65$ & $0.361\pm0.047$ & 2.00 & 0.00 & A \\
8464-3703 & $98.71\pm0.64$ & $190.73\pm5.18$ & - & $13.83\pm0.40$ & $0.117\pm0.078$ & 0.00 & 0.00 & A \\
8137-3703 & $91.22\pm1.25$ & $143.88\pm2.89$ & 0.37 & $6.56\pm2.39$ & $0.154\pm0.048$ & 9.50 & 0.00 & C \\
10216-3703 & $98.51\pm1.50$ & $211.58\pm5.23$ & - & $13.48\pm1.08$ & $0.278\pm0.040$ & 0.00 & 0.00 & A \\
10510-1901 & $100.40\pm0.67$ & $193.53\pm4.37$ & - & $12.19\pm0.67$ & $0.270\pm0.095$ & 2.50 & 0.00 & A \\
11011-1901 & $71.64\pm1.59$ & $232.24\pm5.63$ & 0.57 & $13.09\pm0.35$ & $0.362\pm0.053$ & 0.00 & 0.00 & A \\
11020-1902 & $86.06\pm0.74$ & $211.87\pm2.93$ & 0.48 & $6.74\pm1.85$ & $0.366\pm0.059$ & 0.00 & 7.75 & B \\
11827-1901 & $43.96\pm0.93$ & $165.14\pm2.35$ & 0.48 & $6.42\pm2.48$ & $-0.070\pm0.099$ & 3.50 & 9.25 & B \\
11943-9102 & $90.56\pm1.22$ & $251.36\pm4.47$ & 0.34 & $13.61\pm0.56$ & $0.144\pm0.053$ & 0.00 & 0.00 & A \\
11945-1901 & $100.65\pm1.28$ & $228.63\pm5.01$ & 0.67 & $12.90\pm0.48$ & $0.325\pm0.065$ & 0.00 & 0.00 & A \\
11979-1902 & $108.11\pm1.45$ & $239.61\pm4.42$ & 0.54 & $13.34\pm0.33$ & $0.329\pm0.032$ & 0.00 & 0.00 & A \\
11984-1902 & $102.40\pm0.80$ & $141.08\pm1.93$ & - & $10.33\pm1.99$ & $-0.137\pm0.039$ & 4.50 & 2.00 & B \\
8248-3704 & $87.30\pm0.67$ & $133.57\pm2.76$ & 0.37 & $5.59\pm1.71$ & $-0.006\pm0.064$ & 11.50 & 0.00 & C \\
8710-1902 & $85.89\pm1.45$ & $194.65\pm4.93$ & - & $13.43\pm0.64$ & $0.207\pm0.111$ & 0.00 & 0.50 & A \\
9496-1902 & $94.39\pm1.55$ & $235.65\pm5.81$ & - & $13.09\pm0.65$ & $0.234\pm0.046$ & 0.00 & 5.50 & A \\
9880-1902 & $98.85\pm0.97$ & $175.18\pm4.43$ & 0.29 & $12.82\pm0.66$ & $0.344\pm0.068$ & 0.00 & 0.00 & A \\
11954-1902 & $59.35\pm3.19$ & $135.07\pm7.13$ & 0.82 & $5.87\pm1.33$ & $-0.261\pm0.123$ & 5.00 & 6.20 & B \\
12067-3701 & $85.46\pm1.68$ & $114.56\pm3.28$ & - & $2.96\pm1.94$ & $0.134\pm0.090$ & 12.75 & 0.45 & C \\
7981-1902 & $117.84\pm0.81$ & $164.17\pm1.33$ & 0.77 & $4.13\pm1.52$ & $0.127\pm0.075$ & 11.00 & 1.75 & C \\
11015-1902 & $104.89\pm0.62$ & $192.15\pm3.66$ & 0.29 & $12.47\pm0.93$ & $0.337\pm0.029$ & 0.00 & 2.00 & A \\
11960-1902 & $88.46\pm1.53$ & $272.57\pm6.68$ & 0.42 & $13.26\pm0.72$ & $0.288\pm0.041$ & 0.00 & 5.50 & A \\
11967-3702 & $97.18\pm0.72$ & $197.37\pm5.49$ & 0.29 & $13.56\pm0.33$ & $0.354\pm0.029$ & 0.00 & 0.00 & A \\
11968-1902 & $81.67\pm0.98$ & $204.68\pm4.20$ & 0.62 & $12.72\pm0.49$ & $0.352\pm0.060$ & 0.00 & 2.00 & A \\
11969-1902 & $90.34\pm0.37$ & $187.18\pm4.57$ & 0.44 & $13.21\pm0.81$ & $0.219\pm0.054$ & 0.50 & 0.00 & A \\
12624-3702 & $83.60\pm1.42$ & $218.39\pm4.53$ & 0.45 & $13.01\pm0.63$ & $0.274\pm0.066$ & 0.50 & 2.50 & A \\
12673-1901 & $101.98\pm1.10$ & $256.98\pm4.32$ & 0.52 & $13.28\pm3.15$ & $0.236\pm0.088$ & 0.50 & 0.50 & A \\
\bottomrule
\end{tabular}%
}
\end{table*}

\section{Analysis}\label{sec: analysis}

\begin{figure*}
    \centering
    \includegraphics[width=1\textwidth]{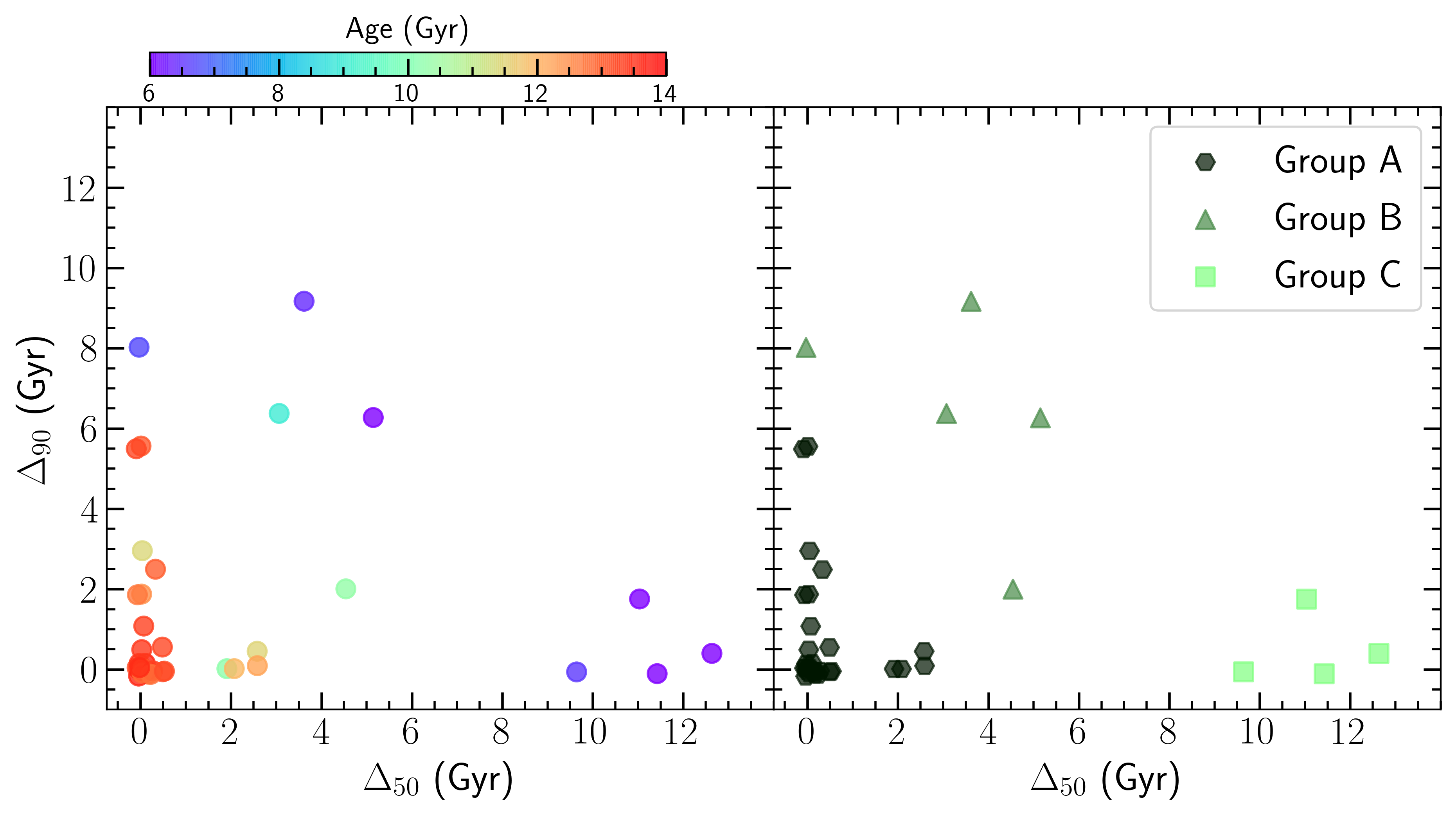}
    \caption{
    Characteristic formation timescales of the compact galaxies in our sample. \textit{Left:} Each galaxy is colored according to its derived mean age measured within 1~$R_e$. The values have been randomly shifted to avoid overlapping points. This figure summarizes when and how fast a galaxy has formed its entire stellar mass. The different locations within this plot will thus characterize different formation pathways.
    \textit{Right: }Same as in left panel, but with galaxies colored according to the results of the $k$-means clustering algorithm (see Section \ref{subsec: clustering}). We label each group as described in the text.}
    \label{fig: A90A50}
\end{figure*}

Combining the mean ages and the formation timescales help us to understand the evolutionary paths of these compact galaxies. The left-side plot of Figure \ref{fig: A90A50} presents the $\Delta_{90}$ and $\Delta_{50}$ values for each galaxy, color-coded by its mean age, as obtained in Section~\ref{sec: methodology}. As we have not applied any regularization in the \ppxf analysis\footnote[3]{Introducing a regularization in the \ppxf analysis can produce a slight differences on the derived values of the stellar populations. However, we have checked that the regularization does not significantly affect the results presented hereafter, as shown in Appendix \ref{app: regularization}.}, the SFHs are bursty, similar to the one presented in Figure \ref{fig: singleSFH}. This makes it more likely for galaxies to have the same $\Delta$ values. We have introduced a small gaussian shift to the in the figure values for illustrative purposes.

The location of the galaxies in Figure \ref{fig: A90A50} can provide information about the different formation channels they have undergone. For example, relic galaxies, which are expected to form very early and extremely fast, almost in a single-like star formation burst, are expected to be located in the lower left corner of Figure \ref{fig: A90A50}, i.e.~with both small $\Delta_{90}$ and $\Delta_{50}$. On the contrary, younger galaxies with more extended SFH will show larger $\Delta_{90}$ and/or $\Delta_{50}$ values.

\subsection{Galaxy clustering via $k$-means} \label{subsec: clustering}
To gain further insight on the formation processes of the compact galaxies in our sample, we have grouped them according to their stellar populations and SFHs. For this, we have used a $k$-means algorithm to classify each of the 37 compact galaxies in three\footnote[4]{According to the elbow method, the optimal number of clusters is 5. However, we have decided to use $k=3$ given the small number of galaxies to be considered. $k=3$ maximizes the differences between groups while keeping enough number of galaxies in each cluster for a reasonable statistical analysis.} different clusters according to their observed properties. The properties considered by the algorithm are: $\Delta_{90}$, $\Delta_{50}$, Age, [M/H], $\Sigma_{1.5}$, and $M_{\star}$. In particular, we wish to focus on the SFH parametrization, therefore we have not introduced any kinematic or size measurements in the clustering algorithm.

We must emphasize that this galaxy allocation in groups only allows to describe the variety of SFHs in our sample. Having only 37 galaxies in our subset prevents from relating these groups with physical galaxy families, with distinct physical properties. Instead, it only allows us to make statements about their different stellar population properties, which are the relevant property for this work.

The mean values of the centroids in each parameter space and the number of galaxies in each group can be found in Table~\ref{tab: centroids}. The algorithm gave more weight to $\Delta_{90}$ and $\Delta_{50}$ in the classification, where the division between groups is more evident. Other stellar population parameters, like metallicity, were used to allocate galaxies with intermediate $\Delta_{90}$ and $\Delta_{50}$ values.

\begin{table*}
\caption{Centroid position for the $k=3$ clusters in the $k$-means classification. Each column represents the position in the respective dimension for the parameters used in the process. In the last column we denote the number of galaxies in each group.}
\label{tab: centroids}
\begin{tabular}{lcccccccr}
\toprule
 & $\boldsymbol{\Delta_{90}}$ & $\boldsymbol{\Delta_{50}}$ & \textbf{Age} & \textbf{[M/H]} & \textbf{[$\boldsymbol{\alpha}$/Fe]} & $\boldsymbol{\log\Sigma_{1.5}}$  & $\boldsymbol{\log(M_{\star}/M_{\odot})}$ & \textbf{\# galaxies} \\
 & \textbf{(Gyr)} & \textbf{(Gyr)} & \textbf{(Gyr)} & \textbf{(dex)} & \textbf{(dex)} & \textbf{(dex)} & \textbf{(dex)} & \\\midrule
Group A & $0.82\pm1.54$ & $0.39\pm0.78$ & $12.89\pm0.78$ & $0.282\pm0.071$ & $0.3 \pm 0.1$ & $10.41\pm0.08$ & $10.51\pm0.24$ & 28 \\
Group B & $6.34\pm2.42$ & $3.20\pm1.75$ & $7.68\pm1.71$ & $0.005\pm0.220$ & $0.1 \pm 0.1$ & $10.51\pm0.16$ & $10.61\pm0.19$ & 5 \\
Group C & $0.55\pm0.72$ & $11.19\pm0.72$ & $4.82\pm1.37$ & $0.102\pm0.063$ & $0.1 \pm 0.1$ & $10.49\pm0.10$ & $10.56\pm0.21$ & 4 \\
\bottomrule
\end{tabular}
\end{table*}

Table \ref{tab: kinematics stell pops} shows the group each galaxy has been allocated into according to this clustering algorithm, which we will refer to as A, B and C. We also show the clustering results in the right plot of Figure~\ref{fig: A90A50}. Similar to the left panel, galaxies are color-coded according to the group they belong.

\begin{figure*}
    \centering
    \includegraphics[width=0.95\textwidth]{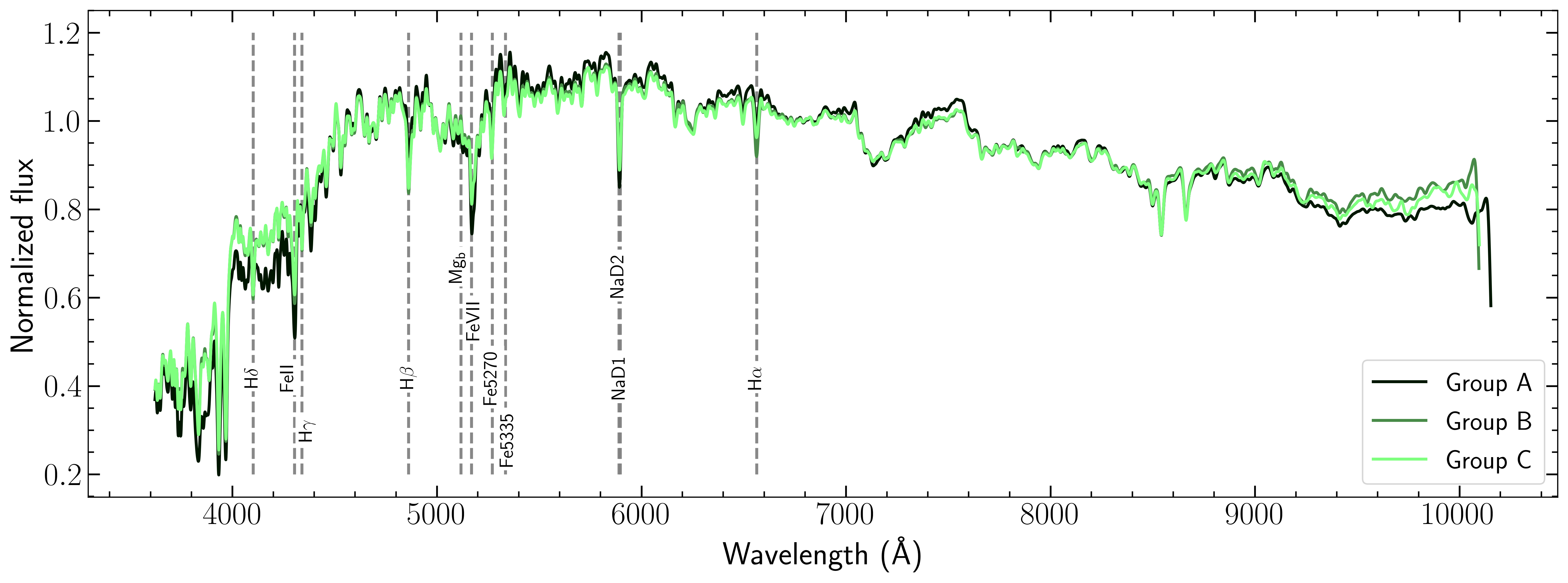}
    \caption{Stacked spectrum of each galaxy group. Relevant spectroscopic lines are also shown, from left to right: H$\alpha$, H$\beta$, H$\delta$, H$\gamma$, Mg$_b$, Fe4303, Fe5159, Fe5270, Fe5335, NaD1 and NaD2.}
    \label{fig: spectra groups}
\end{figure*}

We show in Figure \ref{fig: spectra groups} the stacked spectra of each group of galaxies, along with some relevant spectroscopic lines. It is clear that Group B and C show similar features, while Group A is much different. These behaviors are also seen from the centroid positions in Table \ref{tab: centroids}.

\begin{figure}
    \centering
    \includegraphics[width=0.5\textwidth]{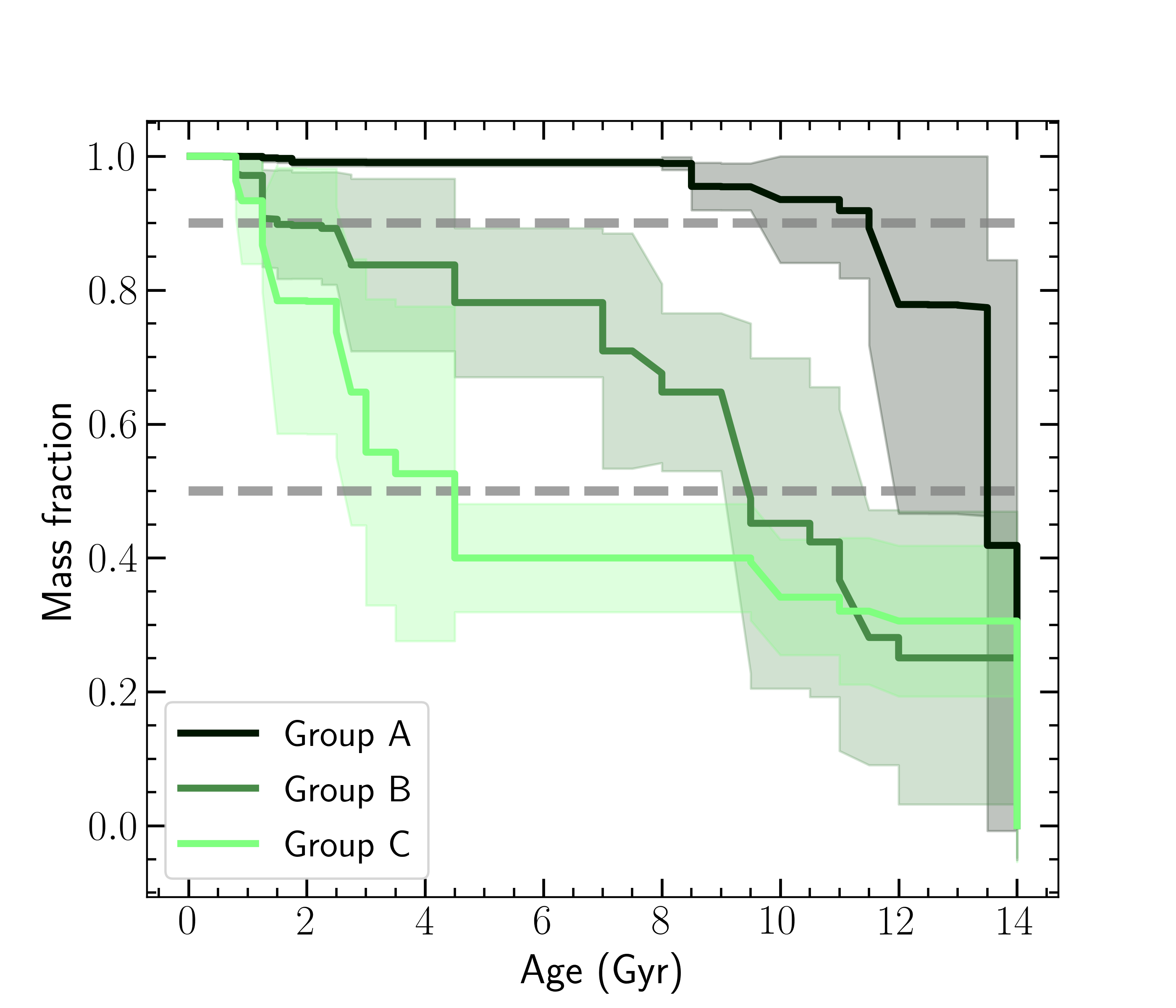}
    \caption{Averaged cumulative SFH for each class of galaxies. The shaded region corresponds to the 1$\sigma$ deviation error. Dotted lines show the position of the 50\% and 90\% of the total mass fraction. Distinctive SFHs are seen for each class.
    }
    \label{fig: mean SFHs}
\end{figure}

Figure \ref{fig: mean SFHs} shows the mean SFH of each group from the $k$-means classification and their 1$\sigma$ errors. As in Figure \ref{fig: singleSFH}, dashed lines show the 50\% and 90\% of the total mass. This figure confirms that the different classes show significant differences in the way they build their stellar mass. These clear differences reinforce the robustness of the $k$-means classification.

According to the behaviors seen in the previous figures and the mean values of the groups (Table \ref{tab: centroids}), the general properties of the three groups of compact galaxies studied in this work are:
\begin{itemize}
    \item Group A: The majority of our compact galaxies, 76\% of the sample, belong to this group. They are old galaxies ($> 12$ Gyr), metal rich ($\sim 0.3$) and with extremely steep SFHs ($\Delta_{90}, \;\Delta_{50} \sim 0$), further supported by their high \alphaFe values (\alphaFe$ = 0.3$ dex). They formed in a single burst-like star formation event. Relic galaxies, if any, would belong to this group. However, this group also includes galaxies with slightly younger ages, of ~$\sim 10$ Gyr, but still with very steep SFHs. These could be `late bloomers', i.e. red nuggets that started their formation at later times. 
    \item Group B: This group includes a $\sim$ 13\% of our compact galaxies. They are intermediate-age galaxies ($\sim 8$ Gyr). They have a wide range of metallicities around the solar-like value ($\sim 0.0 \pm 0.2$ dex), which are consistent with their low \alphaFe values (\alphaFe$ = 0.1$ dex). Their SFHs are extended over time, forming stars until recently. These would be the best candidates for the true low-mass end of ETGs.
    \item Group C: It is the least populated group, with an 11\% of the compact galaxy sample. This group hosts the youngest galaxies ($\sim 5$ Gyr), which show two main star-forming episodes: one at early times ($t\sim 14$ Gyr), which formed roughly the 40\% of their stellar mass, and a later one around $t\sim 4$~Gyr ago lasting until recent times. This is indicated by their high $\Delta_{50}$ values but low $\Delta_{90}$ ones. These galaxies have slightly super-solar metallicities ($\sim 0.1$ dex) and \alphaFe$ = 0.1$ dex. In this case, these could be galaxies that experimented a recent enhancement of their star formations, maybe due to interaction events.
\end{itemize}

\section{Discussion}\label{sec: discussion}
We next compare the properties obtained for the 37 compact galaxies, grouped according to the classification presented in Section \ref{sec: analysis}, to other compact galaxies in the literature. We want to investigate the gap region and its compact galaxies in order to unveil possible relations between compact galaxies at different masses.

We base the following discussion on the parameters derived from the stacked spectra within $1R_e$. Using the values within $1R_e$ also allows to compare our galaxies with other known cEs and CMGs. However, there may be some galaxies for which the parameters derived within $1R_e$ are not fully representative of their behavior. We expect to exploit the spatially-resolved information from MaNGA DR17 IFU observations in future works. 

\subsection{Insights from the stellar populations} \label{subsec: stellar pops}
\begin{figure*}
    \centering
    \includegraphics[width=0.9\textwidth]{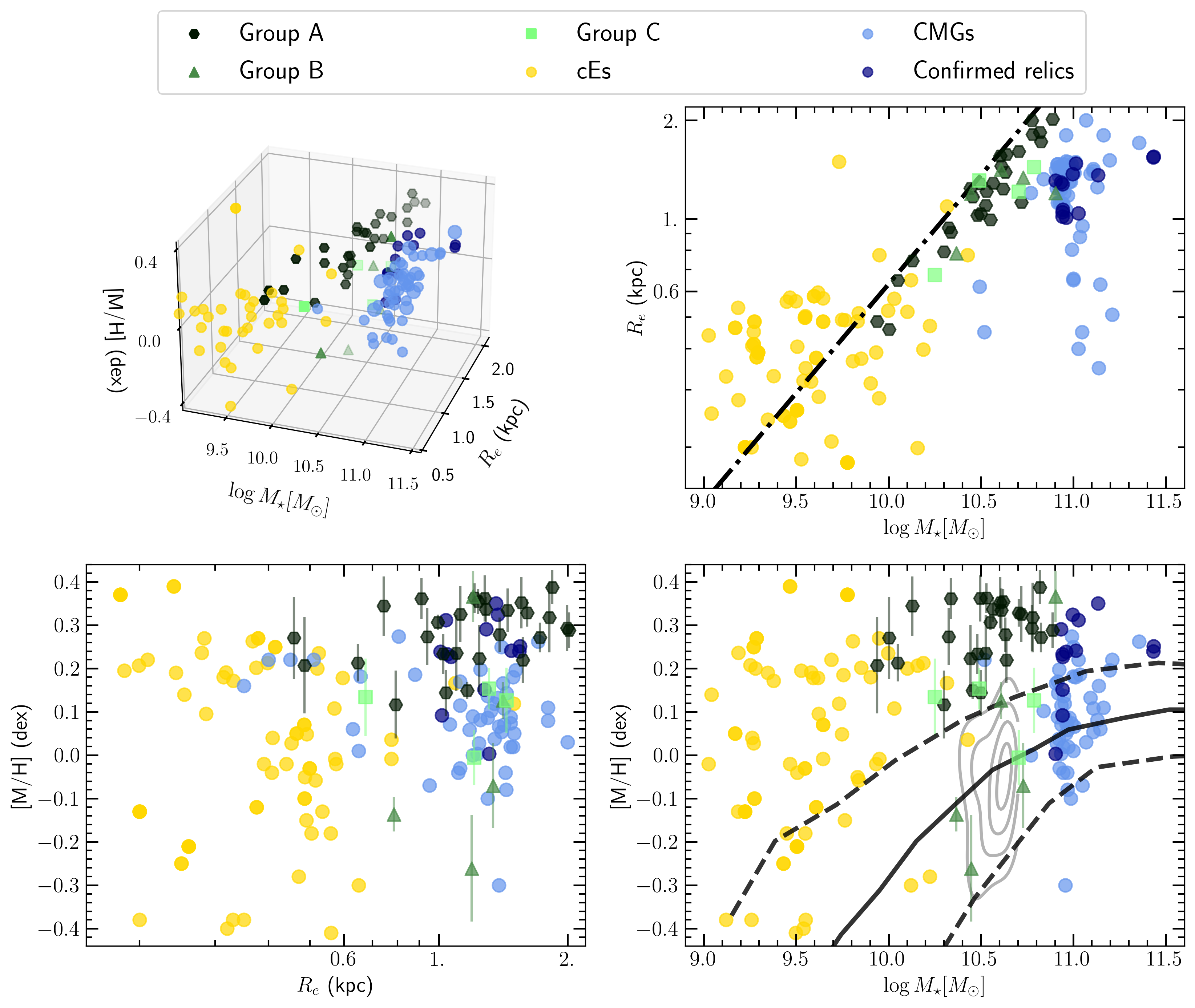}
    \caption{Stellar mass-metallicity-radius fundamental plane. Yellow dots show the values of low-mass cEs from \citet{Janz16, FerreMateu18, FerreMateu21b}. Blue dots show the CMG from \citet{FerreMateu12, FerreMateu15, FerreMateu17, Trujillo14, Yildirim17, Spiniello21b}, where darker blue dots show the those CMG that have been confirmed as relic galaxies. Our compact galaxies are separated according to their classification using the $k$-means algorithm (described in Section \ref{subsec: clustering}). The different projections of the fundamental plane are also shown. The dash-dotted line in the mass-size plot shows the compactness limit adopted in this work, as in Figure \ref{fig:Mass radius}. The mass-metallicity relation at $z\sim 0$ from \citet{Gallazzi21} is shown as a dashed line in the corresponding plot, where the solid line shows the median value of the fitting and the dashed lines the 16\% and 84\% percentiles. The gray contour in this plot shows the position in the plane of non-compact MaNGA ETGs. Our compact galaxies successfully fill the mass gap between cEs and CMGs. Each group shows characteristic metallicities which may be the result of their origins.}
    \label{fig: fundamental plane}
\end{figure*}

In this work we have used the mass-size relation to select the compact galaxies in the MaNGA DR17 sample. However, the mass-metallicity relation (MZR) is one of the most tight relations, whereby the more massive galaxies tend to be more metal rich \citep[e.g.][]{Tremonti04, Gallazzi06, Panter08, Saviane14, Kirby20, Henry21, Langeroodi22}. Figure \ref{fig: fundamental plane} shows the relation between the stellar mass, stellar metallicity and effective radius, along with their projections. In addition to the 37 compact galaxies analyzed in this work, we also show the location of a sample of cEs from \cite{Janz16, FerreMateu18, FerreMateu21b}, and CMGs from \cite{FerreMateu12, FerreMateu15, FerreMateu17, Trujillo14, Yildirim17, Spiniello21b}. 

The top-right projection in Figure \ref{fig: fundamental plane} shows the stellar mass-size relation. This projection was already studied in Section \ref{sec: Sample} (Figure \ref{fig:Mass radius}). We now can confirm that our compact galaxy sample effectively fills the gap between cEs and CMGs, as suggested by \cite{FerreMateu21b}. The bottom-left projection shows the metallicity-size relation, where no clear relation is seen (neither previously known).

The bottom-right panel of Figure \ref{fig: fundamental plane} presents the mass-metallicity relation of compact galaxies, together with the \cite{Gallazzi21} scaling relation for ETGs at $z\sim 0$. This is a crucial projection to better understand the nature of the galaxies. Overall, galaxies that were larger and more massive, but become compact due to stripping events are expected to have higher metallicities than the average scaling relation. On the contrary, those formed in-situ are expected to follow the local scaling MZR. In this projection, cEs present the largest deviations from the MZR, with  the majority of them laying above the MZR \citep{FerreMateu21b}. This is representative of the fact that the majority of cEs are known to be the result of stripping a dwarf or low-mass galaxy. However, there is a small fraction of cEs that are closer to the scaling relation (or within the scatter), which have been proposed to have an intrinsic origin \citep{FerreMateu18, FerreMateu21b, Kim20}. CMGs follow in general the MZR of massive galaxies, with a scatter consistent with the intrinsic one of the relation. Most of the relic galaxies are indeed outliers of the relation. They show higher metallicities than the non-relic CMGs and normally-sized ETGs, probably due to the fact that they missed the second evolutionary phase, which decreases the galaxy total metallicity \citep{FerreMateu17, Spiniello21b}.

As described in \citet{Gallazzi21}, their MZR is calculated based on line indices and a library of parametrized SFHs, which is a different methodology than the one in the present work. We have overplotted in the stellar mass-metallicity plot in Figure \ref{fig: fundamental plane} the density distribution of $\sim$~60 ETGs galaxies with $10.3 < \log M_{\star}/ M_{\odot} < 10.7$ and $3.5 < R_e \text{[kpc]} < 4.5$ in the MaNGA DR17 sample. These ranges of mass and size are considered to be usual among ETGs. The metallicity values for these galaxies were estimated using the same methodology as described in Section~\ref{sec: methodology}. The contour plot reveals that the \citet{Gallazzi21} relation describes well the behavior of non-compact ETGs analyzed with our \ppxf-based methods. Hence, it can be used to analyze our compact galaxy sample as well.


We find that the compact galaxies in this work show a variety of properties in this projection. Group A galaxies are clear outliers to the MZR, being much more metal-rich than the non-compact ETGs. Given their steep SFHs and old stellar populations, some of these galaxies could be good candidates for intermediate-mass relic galaxies. Group C galaxies appear in the limiting region between being outliers and the intrinsic scatter of the MZR. Finally, the MZR is best followed by Group B galaxies. This agrees with their continuous and extended SFH, as the newborn stars should follow the current MZR, suggesting an intrinsic origin.

Interestingly, there is one extreme outlier from Group B in the MZR. This galaxy is also the one with the highest mass in our sample, 11020-1902, which followed the mass-size relation of $z\sim 2$ galaxies in Figure \ref{fig:Mass radius}. In Figure \ref{fig: A90A50} we see that this galaxy is the only in Group B that has $\Delta_{50}\sim 0$ but very large $\Delta_{90}$, with intermediate stellar ages ($\sim 8$ Gyr). The initial hypothesis was that this could be the best candidate for a relic galaxy, but the recovered SFH does not fully support this.

Another possibility to explain the origin of this galaxy is being a `late-bloomer' \citep{FerreMateu12, FerreMateu21b}. These are galaxies that followed the formation pathway of massive galaxies, but that started forming stars much later in cosmic time. `Late-bloomers' will thus have intermediate ages ($\sim$8-10 Gyr) but very short $\Delta_{50}$. If their $\Delta_{90}$ is also small, then these could be the replica of the massive relic galaxies. In fact, we see that only one galaxy in Group A shows small $\Delta_{50},\;\Delta_{90}$ and intermediate ages, 8323-1901. However, if the `late-bloomer' suffers wet interacting processes, these would trigger star-forming events, increasing the value of $\Delta_{90}$, as we see for 11020-1902. All these speculations on the actual nature of individual galaxies will be revisited in future works exploiting MaNGA IFU data.

\begin{figure}
    \centering
    \includegraphics[width=0.49\textwidth]{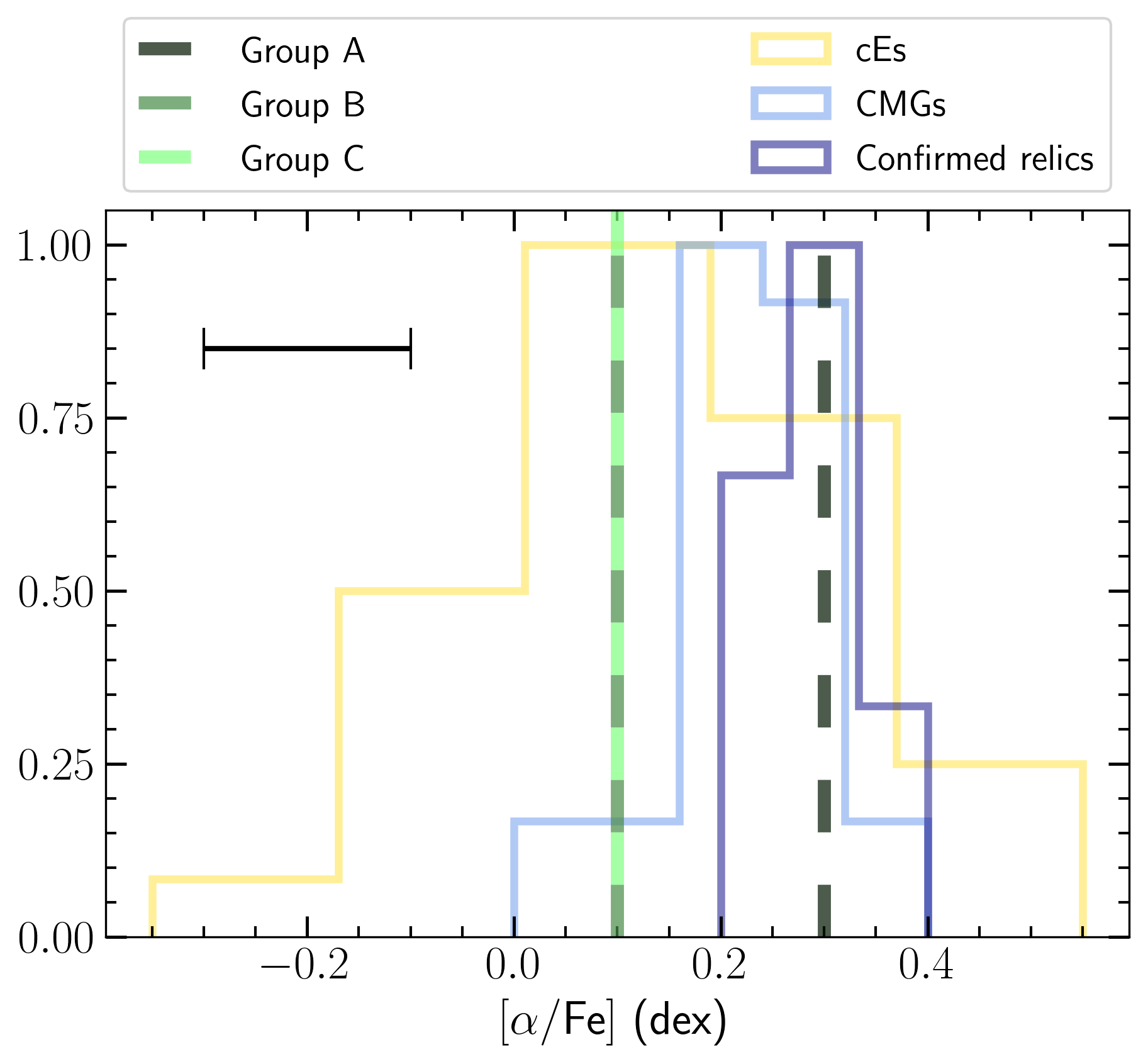}
        \caption{\alphaFe distributions of compact galaxies. Each group histogram is colored as in previous figures: the yellow and blue histograms show the distributions of a sample of cEs \citep{FerreMateu18, FerreMateu21b} and CMGs \citep{Spiniello21a, Trujillo14, Yildirim17, FerreMateu12}, respectively. Confirmed relics \alphaFe distribution is shown in solid lines. Each bar is normalized by the maximum number of counts of that sample. For the three groups studied in this work, the mean values of \alphaFe for each group are shown in dashed lines, with the solid line on the left being their typical uncertainties.}
    \label{fig: alpha}
\end{figure}

Another interesting parameter that can help to gain insight into the formation channels of galaxies is the \alphaFe ratio. This value is deeply related to stellar formation processes. A high \alphaFe ratio is representative of a quick star-forming episode, almost single-burst like, while low \alphaFe values are related to more extended SFHs \citep{Matteucci01timescale, Thomas05, delaRosa11, McDermid15}. We present in Figure \ref{fig: alpha} the \alphaFe distribution values of our compact galaxies, compared with those from the cEs and CMGs in the literature. Group A galaxies show the highest \alphaFe values. Their high \alphaFe values are consistent with their early and steep SFHs (see Section~\ref{sec: analysis}). Such high \alphaFe values are particularly similar to those found in confirmed relics, being slightly higher than general CMGs in some cases. On the other hand, Group B and Group C show lower \alphaFe values, compatible with their more extended SFHs. cEs show the largest dispersion of \alphaFe values, indicative of the mixed origin they have.

\subsection{Insights from the stellar kinematics}\label{subsec: sigma mass}

\begin{figure}
    \centering
    \includegraphics[width=0.45\textwidth]{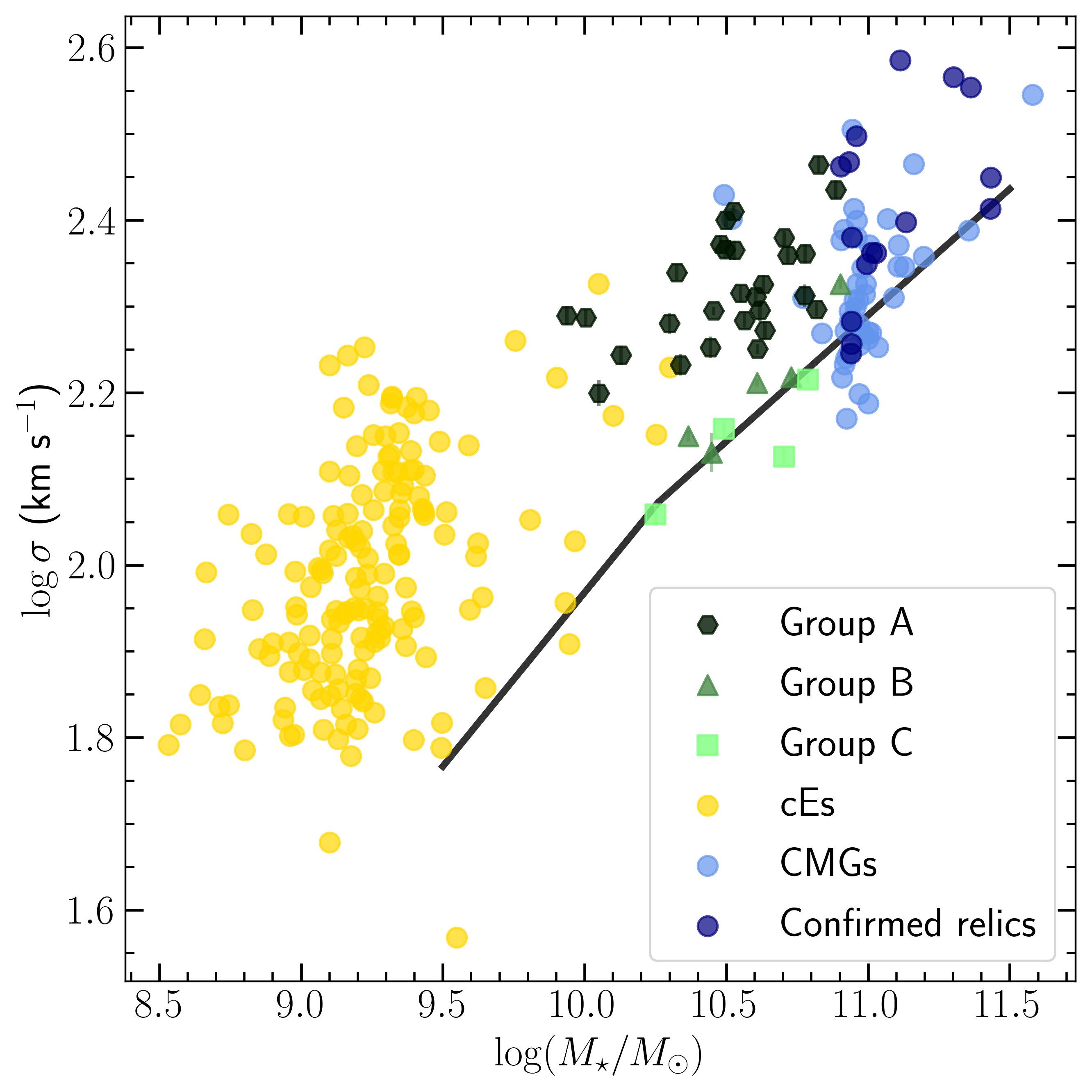}
    \caption{$\sigma$-stellar mass plane for compact galaxies. Our selected 37 cEs are green-colored according to their group belonging. Yellow dots show the position in the plane of the low-mass compact elliptical galaxies from \citet{FerreMateu17, Kim20, FerreMateu21b}. Blue dots show the compact massive galaxies values from \citet{FerreMateu12, Trujillo14, FerreMateu15, FerreMateu17, Yildirim17, Spiniello21b}. Darker blue dots represent the position in the plane of current confirmed relics. The solid black line shows the $\sigma$-stellar mass relation from \citet{Zahid16}.}
    \label{fig: sigma mass}
\end{figure}

In this section we use kinematics to compare the properties of our compact galaxies with those at different masses. One interesting relation is the velocity dispersion-stellar mass relation ($\sigma-M_{\star}$). These two parameters are well correlated as a power law, showing a break at $\log(M_{\star} / M_{\odot}) = 10.26$, which is reported by several authors \citep[e.g.][]{HydeBernardi09, Bernardi11, Cappellari13}. This power law break appears to be similar to the characteristic mass scale at which there is a transition between in-situ to ex-situ processes in compact galaxies \citep[see e.g.,][]{Cappellari16, FerreMateu18, FerreMateu21b, DominguezSanchez20}. This relation has been found to have a narrow intrinsic scatter, despite the break \citep{Evrard08, HydeBernardi09sigma, Sereno15, Zahid16}. Moreover, an important result from the INSPIRE DR1 analysis \citep[][]{Spiniello21b}, and confirmed in INSPIRE DR2 \citep[][]{dAgo23}, is that extreme relics and non-relics behave differently in a $\sigma-M_{\star}$ plot. At a given mass, massive relic galaxies seem to have overall higher stellar velocity dispersion than their non-relic counterparts. And relic galaxies with more extreme SFHs also show higher $\sigma$ values than less extreme ones. Figure~\ref{fig: sigma mass} shows the $\sigma-M_{\star}$ relation for our selected compact galaxies and the cEs and CMGs from the literature, as in previous figures. 

We find that compact galaxies, regardless of the stellar mass, seem to deviate of the $\sigma$-stellar mass relation, in particular at the low-mass end. Only a handful of cEs seem to fit with the \cite{Zahid16} trend. The vast majority of considered cEs show higher velocity dispersion than predicted. However, there is a significant fraction of CMGs that appear to follow the $\sigma - M_{\star}$ relation, although many are still outliers. As found by \cite{Spiniello21b}, extreme relics in the high-mass end are generally CMGs with the highest deviations. 

Regarding the compact galaxies selected in this work, Group A galaxies present the highest velocity dispersions in our sample, being clear outliers of the local scaling relation. They have velocity dispersions similar to higher mass CMGs, and in particular to the confirmed relic galaxies. On the other hand, both Group B and Group C galaxies follow the trend from \cite{Zahid16}. Given their more extended SFHs, this is a further confirmation that these compact galaxies could indeed be the low-mass end of ETGs.

Aiming to investigate the consequences of such high $\sigma$ values for Group A galaxies, we located them in the fundamental plane from \citet{Bernardi20}. In the fundamental plane, the enclosed surface brightness within $1\;R_e$ is related with the stellar velocity dispersion enclosed in the same surface, $\sigma_e$, and $R_e$. This relation is a direct result from the virial theorem, in which the stellar velocity dispersion is related with the mass and the size of the galaxy as $\sigma^2 \sim M_{\star} /R_e$ \citep{Courteau99, Hartl22}. One expects that a virialized system behaviour is well described by the fundamental plane. We show in Figure \ref{fig: fundamental plane} the position of our selected compact galaxies in the MaNGA ETGs fundamental plane from \citet{Bernardi20}. Only a handful of our compact galaxies appear to be described by this fundamental plane. Particularly, Group A galaxies seem to follow an overall different relation, clearly outside the fundamental plane scatter. This would suggest that these galaxies have undergone different formation channels than regular ETGs, such that they would not follow the virial theorem. Due to the small amount of galaxies in Group B and in Group C, we are not able to state whether these groups fortuitously include particular outliers of the plane or these groups are also outliers of the fundamental plane. In any case, understanding what makes these galaxies outliers of that relation would require further investigation, maybe with larger samples of compact galaxies.

\begin{figure}
    \centering
    \includegraphics[width=0.45\textwidth]{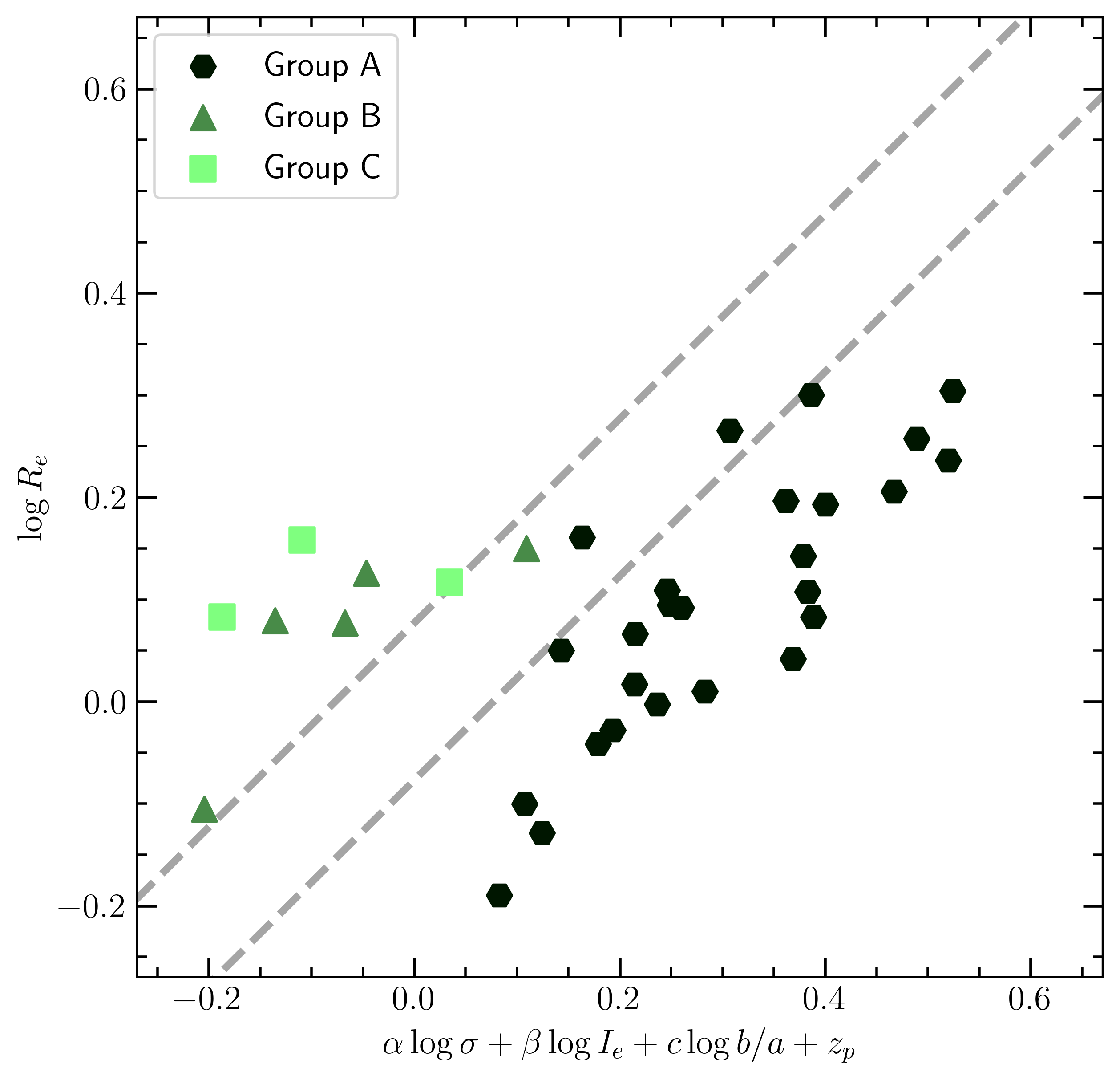}
    \caption{MaNGA ETGs fundamental plane from \citet{Bernardi20}. The plane relates the galaxy effective radius, $R_e$, with the stellar velocity dispersion $\sigma$ and the enclosed surface brightness $I_e$. \citet{Bernardi20} also introduced the semi-axis ratio, $b/a$, as a variable to fit, as well as the free parameter $z_p$. Formerly, kinematics in the fundamental plane are represented as the enclosed stellar velocity dispersion, $\sigma_e$. Due to the small size of our compact galaxies, we have considered $\sigma_e \equiv \sigma$. The plane span is represented by the dashed gray lines (Table 1 from \citet{Bernardi20}, corresponding to the MaNGA ETGs luminosity fundamental plane), which are separated a distance equal to the reported root mean square scatter (rms$_\text{obs}$ = 0.077). Our selected compact galaxies are overlapped with the plane.}
    \label{fig: fundamental plane}
\end{figure}

Finally, another relevant kinematic parameter is the specific angular momentum, $\lambda_R$ (introduced in Equation \ref{eq: lambda}). There appears to be a dichotomy in the kinematics of ETGs: fast rotators (FR) and slow rotators (SR). A fast-rotating galaxy shows a uniform rotational pattern in the innermost regions of their kinematic maps \citep[e.g.][]{Emsellem11, Weijmans14, Foster17, Bilek22}. On the other hand, the kinematic maps of slow-rotating galaxies can show either no rotation or complex features \citep[e.g.][]{Emsellem11, Weijmans14, Foster17}. The specific angular momentum is generally used to distinguish between FR and SR \citep{Emsellem04}. 

\begin{figure}
    \centering
    \includegraphics[width=0.45\textwidth]{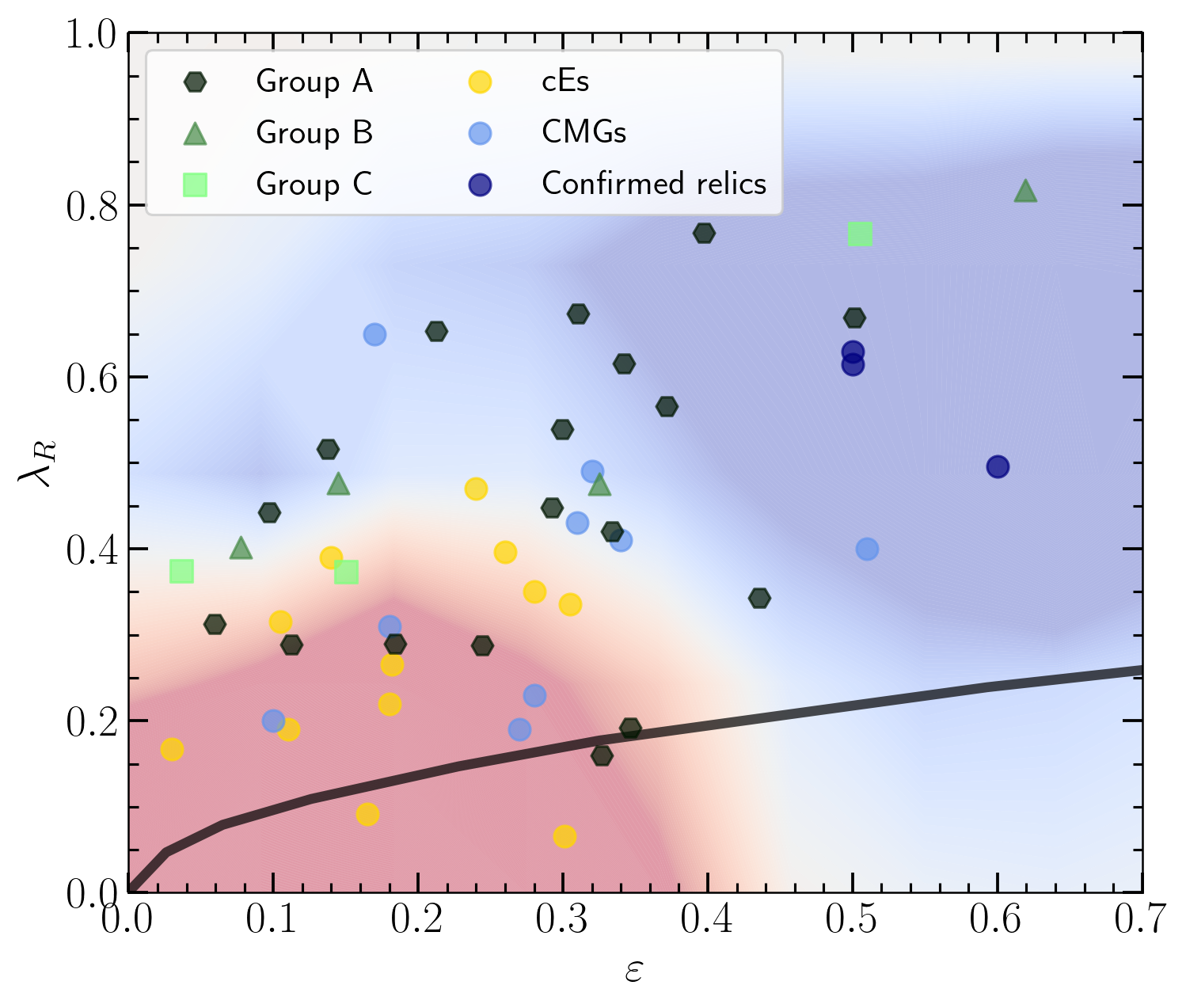}
    \caption{$\lambda_R - \varepsilon$ for the 37 selected compact galaxies. $\lambda_R$ values are corrected for seeing following the methodology in \citet{Graham18}. Literature cEs and CMGs are also shown as in previous figures (when available). Background colors show the density of ETGs (red) or LTGs (blue), according to the values from \citet{Fischer19}. The solid black line represents the \citet{Emsellem04} relation to classify fast-rotating and slow-rotating galaxies. Galaxies below this line are considered SR and above it lie the FR. We find that the majority of compact galaxies, regardless of their mass, are FR.}
    \label{fig: kinematics}
\end{figure}

Figure \ref{fig: kinematics} shows the relation between $\lambda_R$ and the ellipticity of the galaxy ($\varepsilon$) for our 37 compact galaxies. The background is colored to illustrate the density of LTGs (blue) and ETGs (red) galaxies of the whole MaNGA DR17 sample, according to the morphological classification described in Section \ref{sec: Sample}. The $\lambda_R$ values were calculated as described in Section \ref{subsubsection: kinemtatics methodology} and $\varepsilon$ were extracted from the PyMorph VAC. The \cite{Emsellem04} line sets the limit between FR and SR galaxies. As in previous plots, we also show the sample of cEs and CMG for which $\lambda_R$ measurements are available. The $\lambda_R$ parameter was originally conceived to analyse well-resolved galaxy maps. However, our selected compact galxies have a mean effective radius of $\sim1.75$ arcsec. This value is roughly equivalent to 1.5 arcsec, which is the MANGA point-spread-function. Therefore, the analysis concerning $\lambda_R$ remains qualitative due to the lack of proper spatial resolution.

Both CMGs and cEs are, overall, fast rotators, as predicted in simulations from \cite{Naab14}. \cite{FerreMateu21b} observationally checked that cEs tend to be FR. We find that the compact galaxies in our sample are also fast-rotating galaxies\footnote[5]{The FR/SR distinction is performed using $\lambda_R$, which is measured within 1~$R_e$. In any case, rotation would only increase at larger radii. This would not affect the FR classification.}. They all show, however, a wide range of $\lambda_R$ and $\varepsilon$. Those with smaller $\lambda_R$ and smaller $\varepsilon$ show similar distributions to cEs from the literature, while those with larger $\lambda_R$ and $\varepsilon$ kinematics may resemble those of confirmed relics within scatter.

\section{Summary} \label{sec: summary}
Among the diverse family of ETGs, compact galaxies are interesting objects, typically outliers of the local mass-size relation. The compact realm spans over five orders of magnitude in stellar mass. Previous studies have noted that some of their properties, like the stellar populations and kinematics, appear to be related. From this, it is thought that compact galaxies may follow their own scaling relations. However, there was a gap in the mass range between cEs and CMGs, preventing to reach a more firm conclusion. 

In this work, we have analyzed the full MaNGA DR17 sample to find and characterize compact galaxies in this mass region. We have combined the standard mass and size cut criteria with a modified surface mass density threshold to characterize the compactness of the galaxies. Our final sample consists of 37 compact galaxies out of the 10293 galaxies from the MaNGA DR17 dataset. These galaxies seem to follow the mass-size relation at $z \sim 1.5$, despite being local galaxies.

Using stacked spectra up to $1R_e$, we have measured the recessional velocity and the stellar velocity dispersion of each galaxy, along with their stellar population properties, such as age, total metallicity and star formation histories. We find that our selected compact galaxies are all but one fast-rotating galaxies. We define two parameters, $\Delta_{50}$ and $\Delta_{90}$, to characterize the formation timescales of a galaxy. We have then applied a $k$-means algorithm to classify the selected compact galaxies in three different groups, as some galaxies showed clear SFH similarities. The main caveat in this step is that the classification is constrained by the small number of galaxies in the sample.

We have compared our sample to other compact galaxies such as cEs and CMGs, including confirmed relic galaxies. By comparing their main stellar population and kinematic properties, we can suggest different formation pathways for each class. Overall, we find that the main properties shown by each group are:
\begin{itemize}
\item Group A: old galaxies with early and steep SFHs. They were born 14 Gyr ago and have formed all their stars in less than 4 Gyr. They show high mean metallicities and \alphaFe ratio (both with $\sim 0.3$~dex). At a given mass, they show larger velocity dispersions than normal ETGs ($\sigma~\sim~212$~km~s$^{-1}$). Most of them are clear outliers of the current stellar mass-metallicity relation. Therefore, we expect that some of them could be intermediate-mass relics, analogues to those at the high mass end. Moreover, some of the galaxies in this sample could be the so-called `late-bloomers' (i.e. younger relic analogues). 
\item Group B: intermediate-age galaxies ($\sim 8$ Gyr) with continuous SFH over time. Their overall metallicities are lower than those of Group A ($\text{[M/H]}\sim 0$ dex). Given the extended SFHs and the fact that they mostly follow the scaling relations, these galaxies are consistent with the low-mass end of ETGs. Their properties make it unlikely that these galaxies have suffered any interaction with other galaxies, and were probably assembled in-situ. 
\item Group C: young galaxies with a mean age of $\sim 5$ Gyr. Their SFHs reveal an early initial star formation burst, which was then halted in time and resumed $\sim 4$ Gyr ago. A possible explanation is that these galaxies have experienced some recent interaction that drove a cold gas flow into the galaxy center. This could have triggered the late star-forming burst. They show intermediate metallicities ($\text{[M/H]}\sim~0.1$~dex), and the lowest mean \alphaFe ratio ($[\alpha/\text{Fe}]~\sim 0$~dex) among our sample.
\end{itemize}

We have shown that the properties of compact galaxies shift as we consider higher masses. Both $\sigma-M_{\star}$ and $M_{\star}-$[M/H] planes show that in general, cEs are prone to be formed ex-situ. They show unusual high metallicities compared to their stellar mass and some of them also feature high $\sigma$ values. Overall, this suggests a stripping from a larger host galaxy. At the high mass end the number of CMG outliers in the $\sigma-M_{\star}$ and $M_{\star}-$[M/H] relations is lower. It is thus expected that the majority of these have an in-situ origin. The sample of compact galaxies analyzed in this study completely fills the gap between these two families. In fact, they appear to have intermediate properties, which further supports the idea that compact galaxies at different masses are all related.

Even though some of our compact galaxies appear to have similar properties as relic galaxies, we need one more step to reach firm conclusions. We require checking whether the galaxies have undergone any actual changes during its lifetime. To this aim, we expect to take advantage of the spatially-resolved IFU data from MaNGA in future works. With this technology, we would be able to study stellar population gradients, which can reveal more details on the assembly mechanism of these galaxies. We expect to eventually reveal whether a galaxy in the local Universe is still in is pristine stage of a red nugget.

\section*{Acknowledgements}
The authors thank M. Bernardi for his help with the FP and the $\lambda_R$ estimates. The authors also thank the
referee for their constructive and insightful comments that improved the quality of the paper. Part of this work was performed during a JAE fellowship JAEICU-21-ICE-3. AFM acknowledges support from grant CEX2019-000920-S and from RYC2021-031099-I and PID2021-123313NA-I00 of MICIN/AEI/10.13039/501100011033/                           FEDER,UE,NextGenerationEU/PRT. HDS acknowledges support by the PID2020-115098RJ-I00 grant from MCIN/AEI/10.13039/501100011033                            and from the Spanish Ministry of Science and Innovation and the European Union - NextGenerationEU through the Recovery and Resilience Facility project ICTS-MRR-2021-03-CEFCA.

\section*{Data availability}
The MaNGA DR17 data presented are available via the Sloan Digital Sky Survey (SDSS) Science Access Service (SAS): \url{https://data.sdss.org/sas/dr17/manga/}. The different Value Added Catalogs available are listed in \url{https://www.sdss4.org/dr17/data_access/value-added-catalogs/}.

\bibliographystyle{mnras}
\bibliography{bibliography}


\appendix
\section{Effect of the stellar population models in the derived stellar population properties} \label{app: models}

\begin{figure}
    \centering
    \includegraphics[width= 0.45\textwidth]{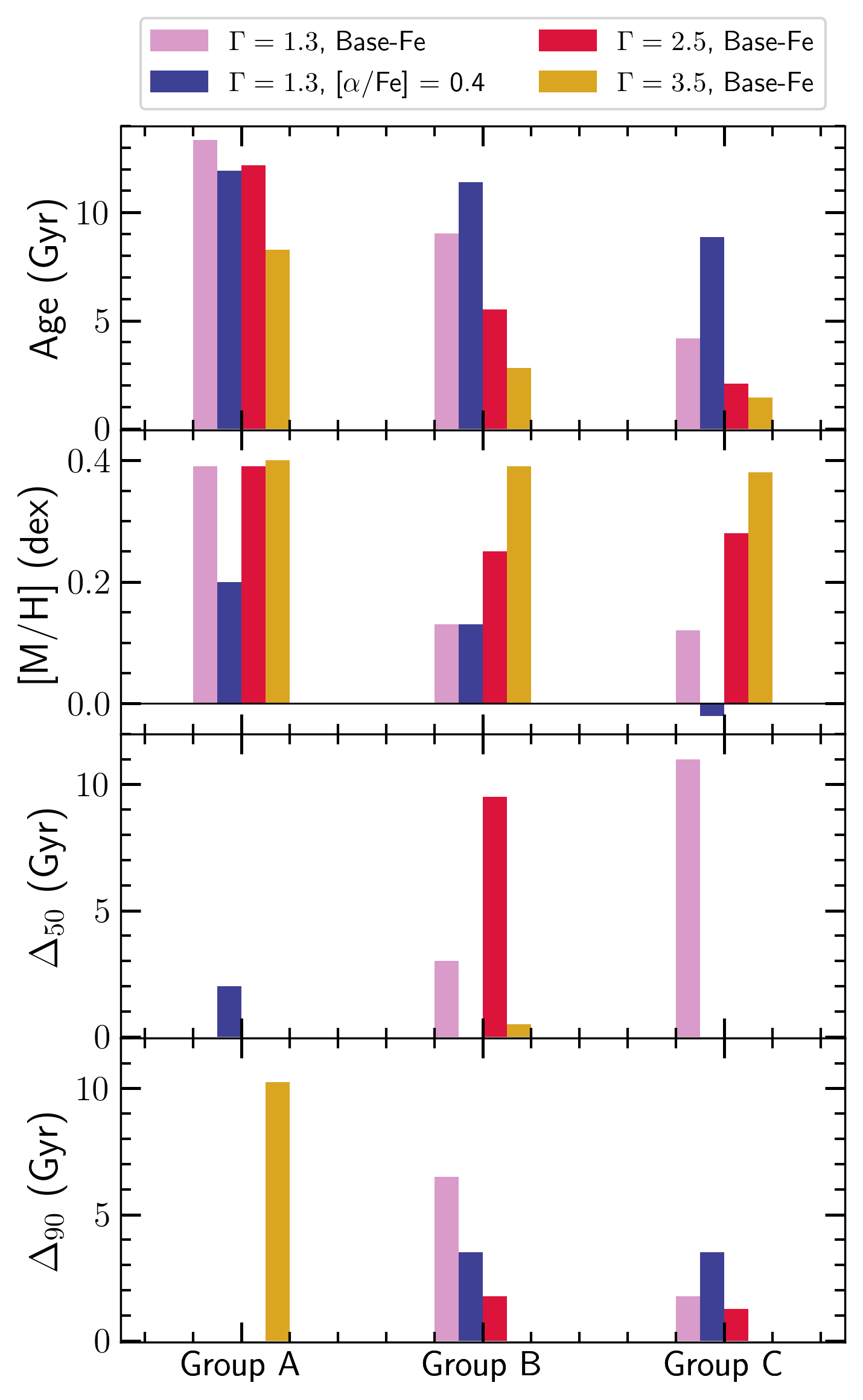}
    \caption{\ppxf-derived mass-weighted age, metallicity, $\Delta_{50}$ and $\Delta_{90}$ values with different stellar population models. The routine was run over the stacked spectrum of a representative galaxy of each group:  9869-1901 (Group A), 8443-1901 (Group B) and 7981-1902 (Group C). The pink bar corresponds to the values extracted with the stellar population models used in the main text. In the $\Delta_{50}$ and $\Delta_{90}$ panels, several bars are not seen because their value is zero.}
    \label{fig: models}
\end{figure}

Aiming to test the robustness of our classification against the use of different stellar populations (SSP) models, we check the impact of changing the IMF and the \alphaFe of the SSP models employed for the fitting routine. This is because we find that all our galaxies are enhanced (see Sect. \ref{subsec: stellar pops}) but also because compact galaxies, in particular if they are relics, have been found to have very steep IMF slopes \citep[e.g][]{FerreMateu17, MartinNavarro23inspire}.

Therefore, we run the \ppxf routine on three galaxies selected as representative of each group: 9869-1901 (Group A), 8443-1901 (Group B) and 7981-1902 (Group C). We use three different sets of SSPs with the same configuration as the described in the main text:
\begin{itemize}
    \item SSP models with $\Gamma = 1.30$ and \alphaFe$ = 0.40$ 
    \item SSP models with $\Gamma = 2.50$ and Base-Fe
    \item SSP models with $\Gamma = 3.50$ and Base-Fe
\end{itemize}

Figure \ref{fig: models} shows the values for the mass-weighted age, metallicity, $\Delta_{50}$ and $\Delta_{90}$ of each group, employing the different SSP models described. For comparison, we also show the values from the main text ($\Gamma = 1.30$ and Base-Fe, see Table \ref{tab: kinematics stell pops}).

We see in Figure \ref{fig: models} that an \alphaFe-enhanced model slightly decreases the age of the Group A galaxy, although it maintains its $\sim$12 Gyr. Its metallicity also decreases. 
However, the derived SFH of the Group A galaxy is still peaked, as seen from its $\Delta_{50}$ and $\Delta_{90}$ parameters. On the other hand, the \alphaFe-enhanced models increase the age values of both Group B and Group C galaxies, although it does not significantly change their metallicities. 
Both the $\Delta_{50}$ and $\Delta_{90}$ values for the Group B and Group C are kept similar compared to the nominal values. However, they are systematically always different than those from Group A. Hence, a clustering classification based on \alphaFe-enhanced models would not differ from that in the main text.

Considering the effect of the IMF on the derived properties, we note that all derived ages decrease with the steepness of the models IMF due to the addition of high-mass stars to the model \citep[e.g][]{FerreMateu13}. These stars lower the overall age of the galaxy and they also increase the overall metallicity. This is particularly seen in the Group B and Group C galaxies. 
Related to the derived SFHs, a $\Gamma = 2.50$ does not change it for the Group A galaxy, i.e., it is still peaked. While a $\Gamma = 3.50$ 
keeps a fast formation for this galaxy, it produces a slightly more extended SFH. However, it has been seen the steep IMF slopes found for compact galaxies are not as extreme as this but are better described by $\Gamma\sim 2.5$ \citep[e.g][]{FerreMateu17, MartinNavarro23inspire}. 
Finally, for Group B and Group C galaxies, increasing the IMF slope clearly changes the shape of their SFH. However, this would not affect the overall allocation in groups, as age is an additional parameter taken into account in the classification. For example, the Group C galaxy SFH becomes single-burst-like when using $\Gamma = 3.50$, but its age is around $\sim$2~Gyr. Therefore it would still not be allocated to Group A.

Therefore we find that our results are consistent against different SSP models, giving a similar classification to that in  Section~ \ref{subsec: clustering}.

\section{Effect of the regularization parameter on the derived stellar population properties}\label{app: regularization}

In this appendix we discuss whether applying a regularization on the \ppxf routine can have an impact on the $k$-means classification. Regularization changes the model weights distribution, and thus it can potentially change the shape of the derived SFH. This, in turn, has a direct impact onto the $\Delta_{50}$, $\Delta_{90}$, mean ages and metallicities recovered. 

We worked with the regularization error, \texttt{regul\_err}, which inverse was then passed to \ppxf as the \texttt{regul} parameter. 

\begin{figure}
    \centering
    \includegraphics[width = 0.4\textwidth]{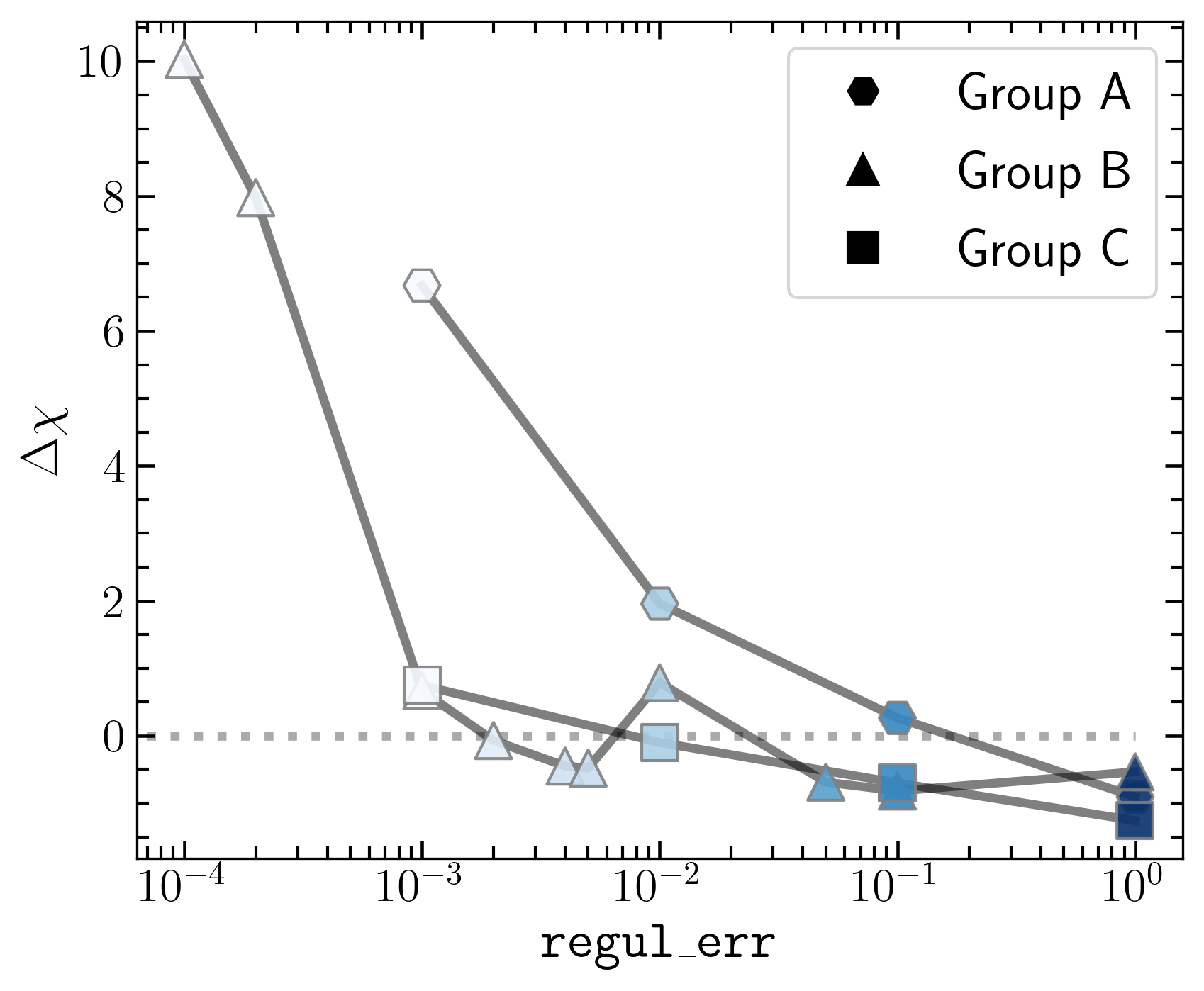}
    \caption{Difference between the desired $\chi^2$ and the actual $\chi^2$ of the fit for a given \texttt{regul\_err}. Each line shows the results from a given galaxy of each group: 9869-1901 for Group A, 8443-1901 for Group B and 7981-1902 for Group C. Each marker is colored according to an increasing \texttt{regul\_err} value.}
    \label{fig: regul chi}
\end{figure}

We expect that any variations due to the regularization will affect all galaxies in each group similarly. Therefore, we use a representative galaxy of each group (the same as in Appendix \ref{app: models} for this analysis. We repeat the \ppxf fitting routine for each galaxy with different regularization errors. According to \cite{Cappellari12ppxf}, the goodness of the fit is characterized by its $\chi^2$ value. In Figure \ref{fig: regul chi} we show the difference between the desired $\chi^2$ and the actual $\chi^2$, represented as $\Delta\chi$, for a varying \texttt{regul\_err}. The best \texttt{regul\_err} value will be that for which $\Delta\chi = 0$. 

Group A galaxy requires almost no regularization ($\texttt{regul\_err}\gtrsim 10^{-1}$), reinforcing our finding that they indeed form quickly. On the other hand, Group B and Group C galaxies require \texttt{regul\_err} around $10^{-2}$. In order to better see the shape of the Group B curve, it was sampled with more \texttt{regul\_err} values.

\begin{figure}
\centering
\includegraphics[width=0.4\textwidth]{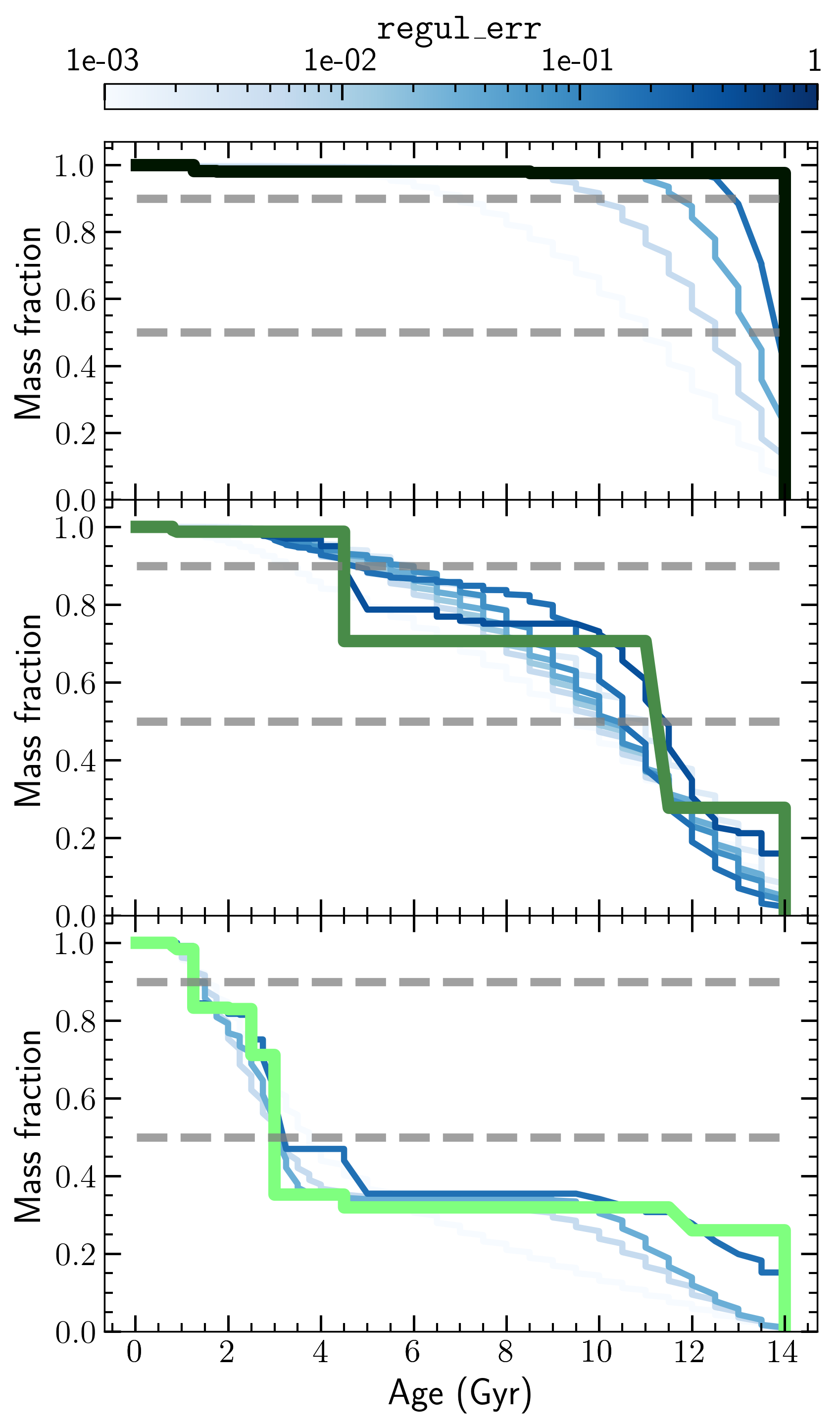}
\caption{Effect of changing the \texttt{regul\_err} parameter on the derived SFHs. From top to bottom: 9869-1901 (Group A), 8443-1901 (Group B) and 7981-1902 (Group C). The non-regularized SFH for a representative galaxy of each group is shown in green, while colored lines correspond to the regularized solutions, as in Figure \ref{fig: regul chi}. The gray dashed lines show the 50\% and 90\% of the total mass fraction. Even though applying a regularization changes the derived SFH, the values of $\Delta_{50}$ and $\Delta_{90}$ do not significantly change.}
\label{fig: regularization variations}
\end{figure}

The derived SFHs for each \texttt{regul\_err} parameter are shown in Figure \ref{fig: regularization variations}, compared to the non-regularized SFH used in the main text. From each panel, it can be seen that the values of $\Delta_{50}$ and $\Delta_{90}$ do not differ significantly regardless of \texttt{regul\_err}. Specially focusing on the SFH with the optimal \texttt{regul\_err} value, we see that Group A galaxy still has $\Delta_{50} = 0$ and $\Delta_{90}$ changes to 2 Gyr. For the Group B galaxy $\Delta_{50}$ increases 1 Gyr and $\Delta_{90}$ decreases 1 Gyr. And for the galaxy in Group C there is no noticeable variation in $\Delta_{50}$ nor $\Delta_{90}$. 
 
Although the $k$-means method in Section \ref{subsec: clustering} allocates galaxies mainly based on their $\Delta_{50}$ and $\Delta_{90}$ values, we consider that there would not be significant differences if regularized solutions had been considered. 

\bsp	
\label{lastpage}
\end{document}